\documentclass{elsart}

\usepackage[square,comma]{natbib}
\usepackage{graphicx}
\usepackage{pxfonts}
\usepackage{lineno}

\usepackage{amssymb}

\journal{}

\begin{document}

\thispagestyle{empty}
\begin{Large}
\textbf{DEUTSCHES ELEKTRONEN-SYNCHROTRON}

\textbf{\large{Ein Forschungszentrum der Helmholtz-Gemeinschaft}\\}
\end{Large}

DESY 12-034

February 2012

\begin{eqnarray}
\nonumber &&\cr \nonumber && \cr \nonumber &&\cr
\end{eqnarray}
\begin{eqnarray}
\nonumber
\end{eqnarray}
\begin{center}
\begin{Large}
\textbf{Self-seeding scheme for the soft X-ray line at the European
XFEL}
\end{Large}
\begin{eqnarray}
\nonumber &&\cr \nonumber && \cr
\end{eqnarray}

\begin{large}
Gianluca Geloni,
\end{large}
\textsl{\\European XFEL GmbH, Hamburg}
\begin{large}

Vitali Kocharyan and Evgeni Saldin
\end{large}
\textsl{\\Deutsches Elektronen-Synchrotron DESY, Hamburg}
\begin{eqnarray}
\nonumber
\end{eqnarray}
\begin{eqnarray}
\nonumber
\end{eqnarray}
ISSN 0418-9833
\begin{eqnarray}
\nonumber
\end{eqnarray}
\begin{large}
\textbf{NOTKESTRASSE 85 - 22607 HAMBURG}
\end{large}
\end{center}
\clearpage
\newpage

\begin{frontmatter}



\title{Self-seeding scheme for the soft X-ray line at the European XFEL}


\author[XFEL]{Gianluca Geloni\thanksref{corr},}
\thanks[corr]{Corresponding Author. E-mail address: gianluca.geloni@xfel.eu}
\author[DESY]{Vitali Kocharyan}
\author[DESY]{and Evgeni Saldin}

\address[XFEL]{European XFEL GmbH, Hamburg, Germany}
\address[DESY]{Deutsches Elektronen-Synchrotron (DESY), Hamburg,
Germany}

\begin{abstract}
This paper discusses the potential for enhancing the capabilities of
the European FEL in the soft X-ray regime. A high longitudinal
coherence will be the key to such performance upgrade. In order to
reach this goal we study a very compact soft X-ray self-seeding
scheme originally designed at SLAC \cite{FENG,FENG2}. The scheme is
based on a grating monochromator, and can be straightforwardly
installed in the SASE3 undulator beamline at the European XFEL. For
the European XFEL fully-coherent soft X-ray pulses are particularly
valuable since they naturally support the extraction of more FEL
power than at saturation by exploiting tapering in the tunable-gap
SASE3 undulator. Tapering consists of a stepwise change of the
undulator gap from segment to segment. Based on start-to-end
simulations we show that soft X-ray FEL power reaches about $800$
GW, that is about an order of magnitude higher than the SASE level
at saturation ($100$ GW). The self-seeding setup studied in this
work is extremely compact (about $5$ m long), and cost-effective.
This last characteristic may justify to consider it as a possible
addition to the European XFEL capabilities from the very beginning
of the operation phase.
\end{abstract}

%
%
%
\end{frontmatter}



\section{\label{sec:intro} Introduction}

The quality of the output radiation of X-ray SASE FELs is far from
ideal. SASE X-ray beams are characterized by nearly full transverse
coherence but only limited longitudinal coherence
\cite{LCLS2}-\cite{tdr-2006}. However, many experiments require both
transverse and longitudinal coherence. In principle, one can create
a longitudinal coherent source by using a monochromator located in
experimental hall, but this is often undesirable because of
intensity losses. An important goal for any advanced XFEL facility
is the production of X-ray radiation pulses with minimum allowed
photon energy width for a given pulse length, that is
Fourier-limited pulses. In this way, no monochromator is needed in
the experimental hall. Self-seeding is a promising approach to
significantly narrow the SASE bandwidth and to produce nearly
transform-limited pulses \cite{SELF}-\cite{WUFEL2}. Considerable
effort has been invested in theoretical investigation and R$\&$D at
LCLS leading to this capability \cite{CDRL2}.

In general, a self-seeding setup consists of two undulators
separated by a photon monochromator and an electron bypass, normally
a four-dipole chicane. The two undulators are resonant to the same
radiation wavelength. The SASE radiation generated by the first
undulator passes through the narrow-band monochromator. A
transform-limited pulse is created, which is used as a coherent seed
in the second undulator. Chromatic dispersion effects in the bypass
chicane smear out the microbunching in the electron bunch, produced
by the SASE lasing in the first undulator. The electrons and the
monochromatized photon beam are recombined at the entrance of the
second undulator, and radiation is amplified by the electron bunch
until saturation is reached. The required seed power at the
beginning of the second undulator must dominate over the shot noise
power within the gain bandpass, which is order of a kW in the soft
X-ray range.

For soft X-ray self-seeding, a monochromator usually consists of a
grating \cite{SELF}. Recently, a very compact soft X-ray
self-seeding scheme has appeared, based on a grating monochromator
\cite{FENG,FENG2}. The proposed monochromator is composed of only
three mirrors and a rotational VLS grating. A preliminary design of
the gratings adopts a constant focal-point mode in order to have
fixed slit location. When tuning the photon energy in the range
between $250$ and $1000$ eV, the variation in the optical delay is
limited to $10 \%$ of the nominal value of $2.5$ ps.  A short
magnetic chicane delays the electron bunch accordingly, so that the
photon beam passing through the monochromator system recombines with
the same electron bunch. The chicane provides a dispersion strength
of about $2$ mm in order to match the optical delay and also smears
out the SASE microbunching generated in the first undulator.

In this article we study the performance of the above-described
scheme for the European XFEL upgrade. The installation of the
chicane does not perturb the undulator focusing system and allows
for a safe return to the baseline mode of operation. The inclusion
of a chicane is not expensive and may find many other applications
\cite{MEAS,PUMP}.

With the radiation beam monochromatized down to the Fourier
transform limit, a variety of very different techniques leading to
further improvement of the X-ray FEL performance become feasible. In
particular, the most promising way to extract more FEL power than
that at saturation is by tapering the magnetic field of the
undulator \cite{TAP1}-\cite{TAP4}. A significant increase in power
is achievable by starting the FEL process from monochromatic seed
rather than from noise, \cite{FAWL}-\cite{WANG}. In this paper we
propose a study of the performance of the soft X-ray self-seeding
scheme for the European XFEL, based on start-to-end simulations for
an electron beam with 0.1 nC charge \cite{S2ER}. Simulations show
that the FEL power of the transform-limited soft X-ray pulses may be
increased up to 800 GW by properly tapering the baseline undulator.
In particular, it is possible to create a source capable of
delivering fully-coherent, $10$ fs (FWHM) soft X-ray pulses with
$0.6 \cdot 10^{14}$ photons per pulse at the wavelength of 1.5 nm.

\section{Possible self-seeding scheme with grating monochromator for the baseline undulator
SASE3}

\begin{figure}[tb]
\includegraphics[width=1.0\textwidth]{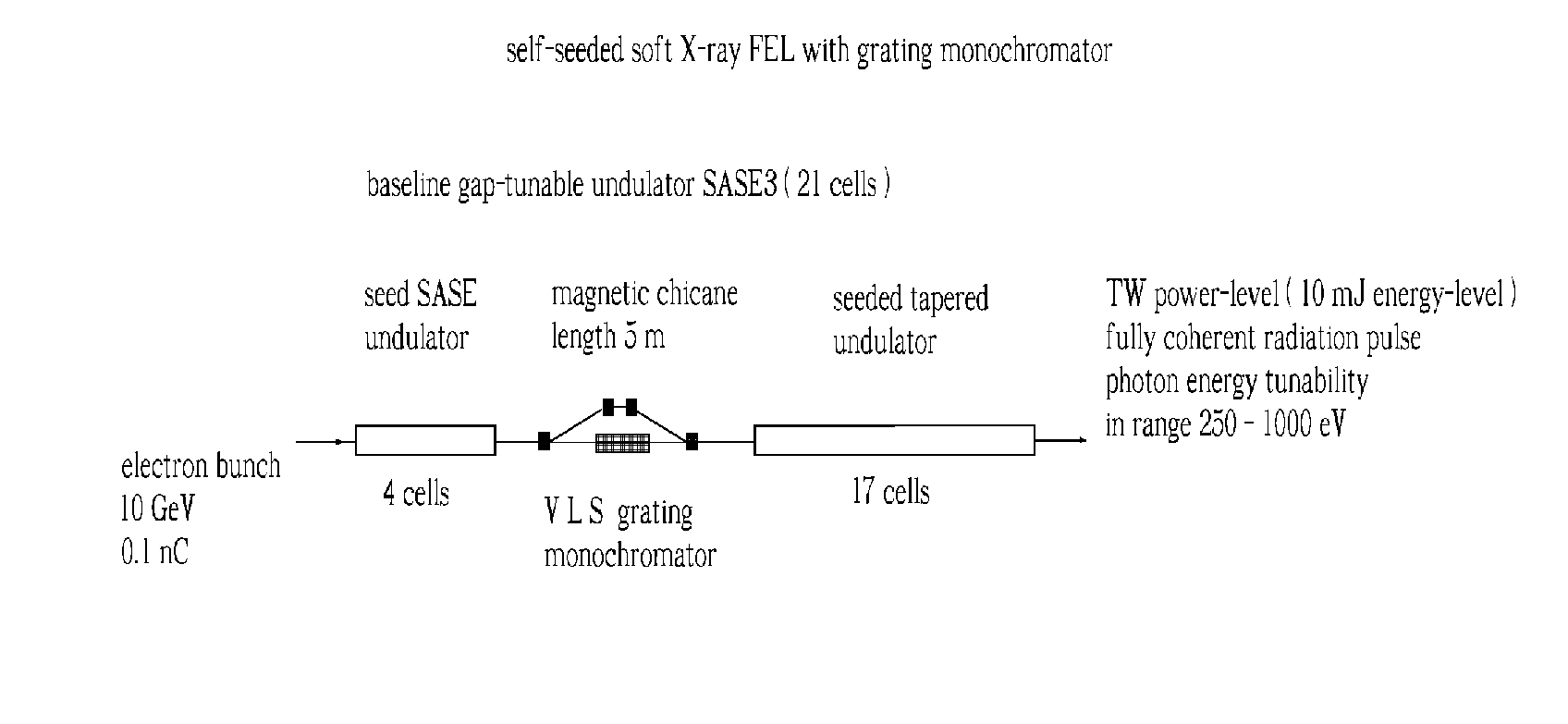}
\caption{Design of the self-seeding setup based on the European XFEL
baseline undulator system for generating highly monochromatic, high
power soft X-ray pulses. The method exploits a combination of a
self-seeding scheme with a grating monochromator and of a undulator
tapering technique. The self-seeding setup is composed of a compact
(about $2.5$ ps optical delay) grating monochromator originally
proposed at SLAC \cite{FENG,FENG2} and a 5 m-long magnetic chicane.
The magnetic chicane accomplishes three tasks by itself. It creates
an offset for monochromator installation, it removes the electron
microbunching produced in the upstream (seed) undulator, and it acts
as a electron beam delay line for monochromator optical delay
compensation.} \label{soft1}
\end{figure}

\begin{figure}[tb]
\includegraphics[width=1.0\textwidth]{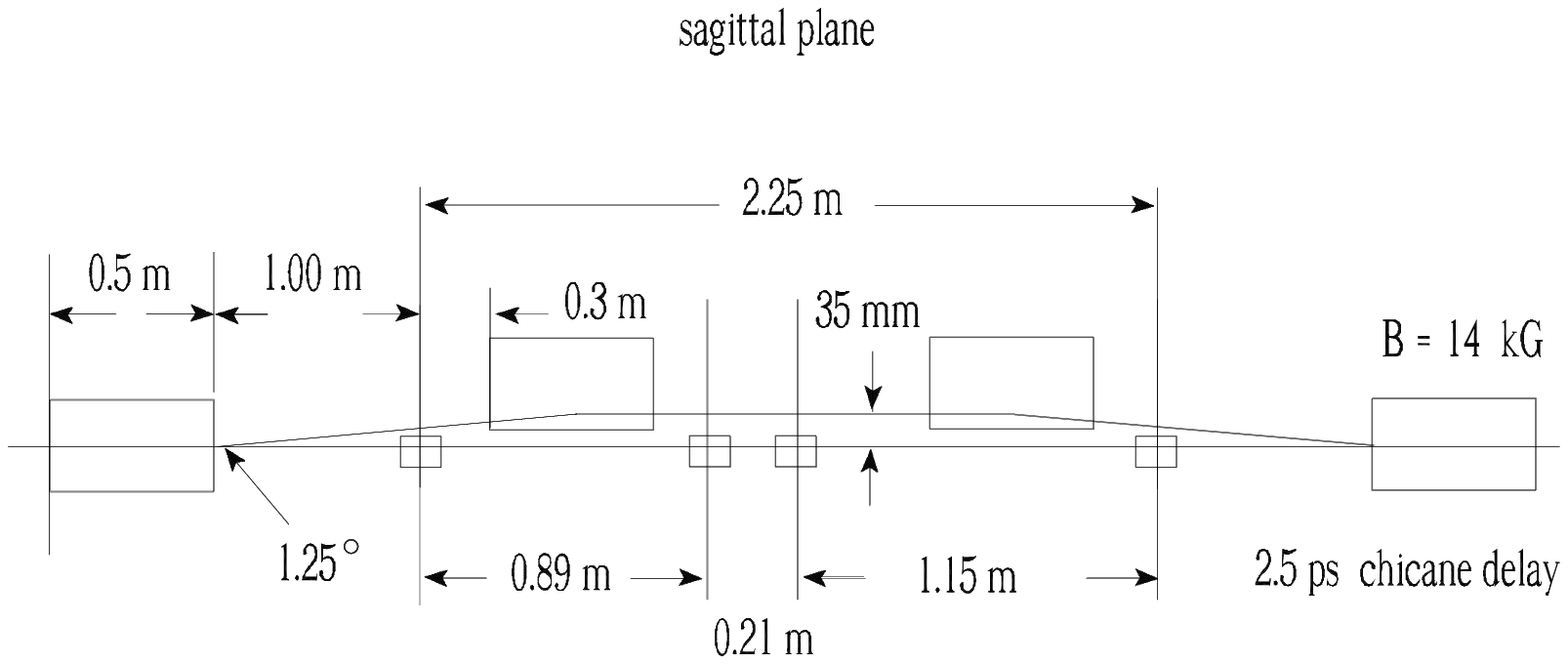}
\caption{Plan view of the self-seeding setup with compact grating
monochromator originally proposed at SLAC \cite{FENG,FENG2}.}
\label{soft2}
\end{figure}

\begin{figure}[tb]
\includegraphics[width=1.0\textwidth]{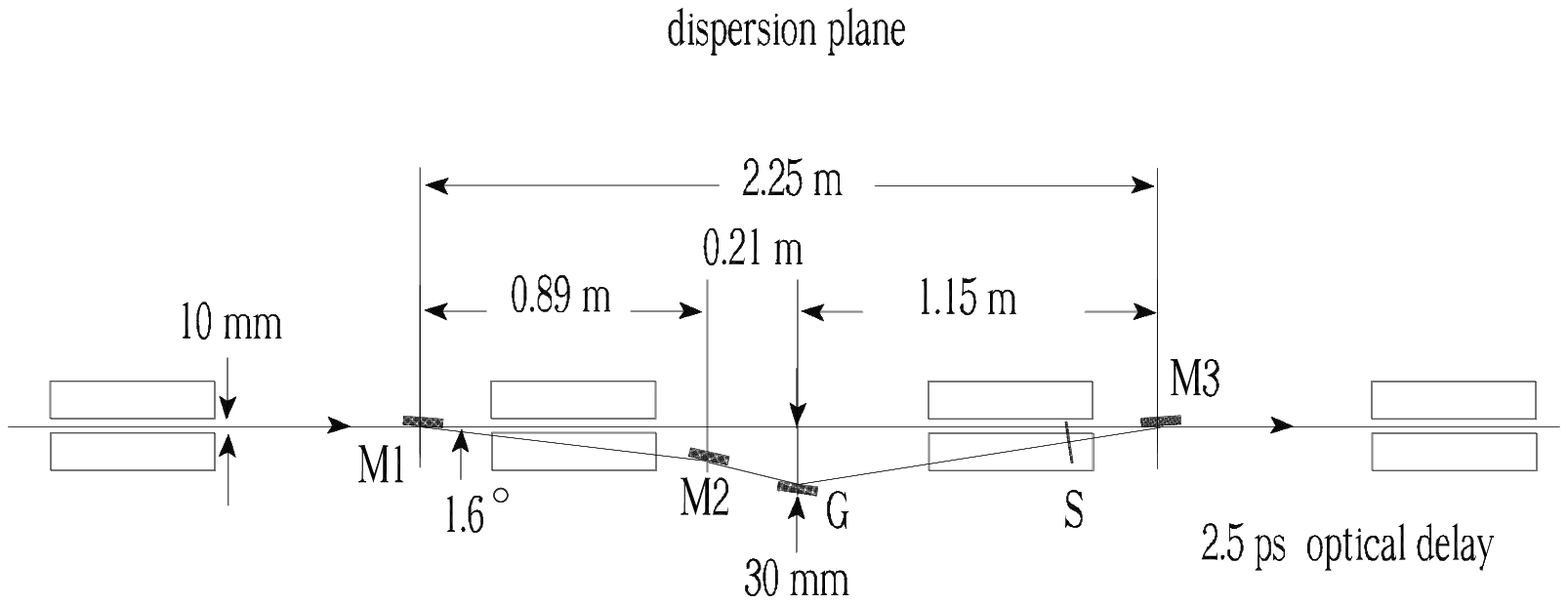}
\caption{Elevation view of the self-seeding setup with compact
grating monochromator originally proposed at SLAC
\cite{FENG,FENG2}.} \label{soft3}
\end{figure}

\begin{figure}[tb]
\includegraphics[width=1.0\textwidth]{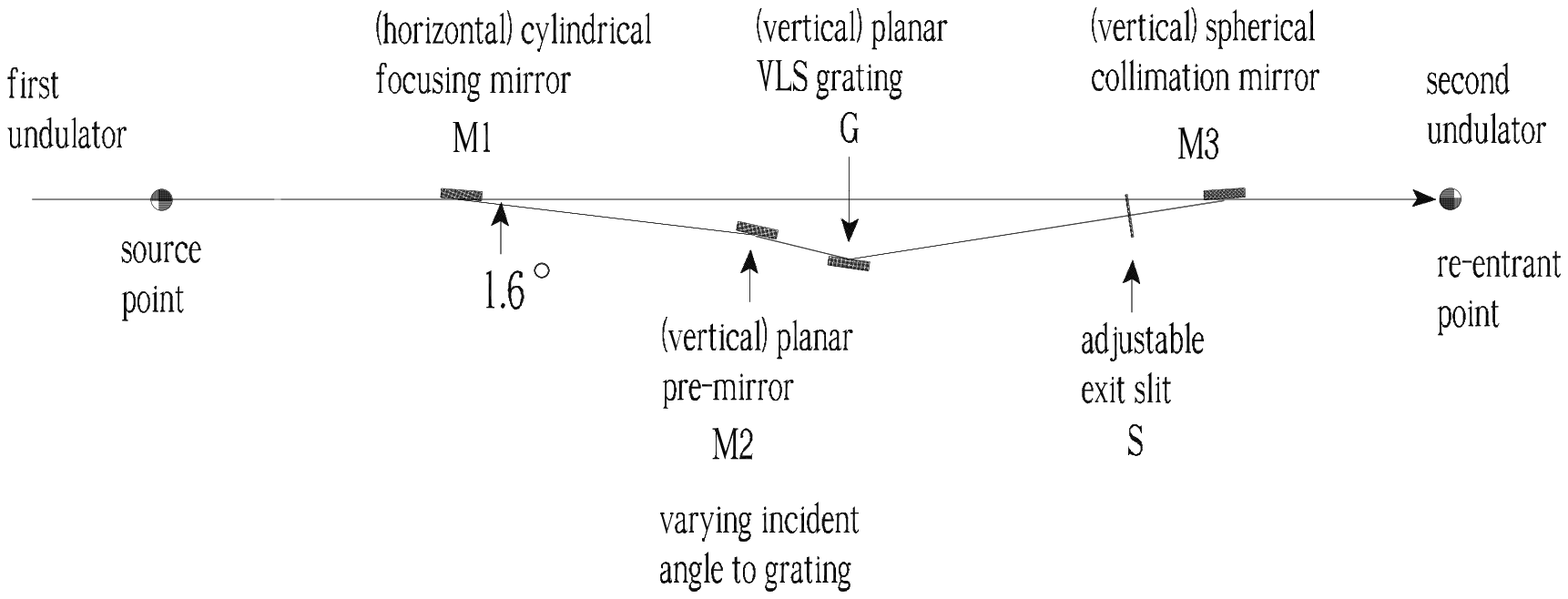}
\caption{Optics for the compact grating monochromator originally
proposed at SLAC \cite{FENG,FENG2} for the soft X-ray self-seeding
setup.} \label{soft4}
\end{figure}
Any self-seeding setup should be compact enough to fit one undulator
segment. The elements of the adopted electron bypass design are four
$0.5$ m-long dipole magnets of rectangular shape. Under the
constraints imposed by space and strength of the magnetic field it
is only possible to operate at an electron beam energy of $10$ GeV.
The first dipole deflects the beam by $1.25$ degrees. After $1.8$ m,
the second dipole deflects  the beam back in a direction parallel to
the straight beampath  at a distance of about 3 cm, which is
sufficient for the installation of the optical elements of the
monochromator. The total elongation of the electron beampath is
approximately $1$ mm. The layout of the bypass is shown in Fig.
\ref{soft2} and Fig. \ref{soft3}.

The monochromator design should be compact enough to fit with this
magnetic chicane design. In particular, the optical delay should be
matched to that induced by the magnetic chicane on the electron
beam. The design adopted in this paper is the novel one by Y. Feng
et al. \cite{FENG,FENG2}, and is based on a  planar VLS grating. It
is equipped only with an exit slit. Such design includes four
optical elements, a cylindrical and spherical focusing mirrors, a
VLS grating and a plane mirror in front of the grating. The
specifications of the monochromator are summarized below:

\begin{itemize}
\item The total optical beamline length, from the source to the image,
is about 6 m.

\item The total length between first and last optical components is about
2.3 m

\item The delay of the photons is about  $2.5$ ps.

\item The monochromator is continuously tunable in the photon energy range
$250-1000$ eV.

\item The resolving power is about $4500$ at the wavelength around $1.5$ nm.

\item The photon beam size and divergence at the entrance of the 2nd
undulator are close to those at the exit of the first undulator.

\item The transmission of the beamline is close to $10 \%$ for wavelengths around
$1.5$ nm.
\end{itemize}
The optical layout of the monochromator is schematically shown in
Fig. \ref{soft4}. The first optical component in the monochromator,
$M1$, is a cylindrical mirror which focuses the input photon beam in
the sagittal plane at the entrance point, and deflects it in the
dispersion plane of about $1.6$ degree. This mirror has a long focal
length of about $5$ m. The last optical component, $M3$, is a
collimation mirror that refocuses the photon beam at the entrance
point. It introduces a deflection angle in the dispersion plane of
about $1.2^\circ$. This mirror has a short focal distance of about
$20$ cm. The second mirror, $M2$, is plane mirror set in front of
the grating. The function of this mirror is to illuminate the
grating at the proper angle of incidence, which is chosen so that
the correct wavelength is approximately focused at the exit slit.
The monochromator scanning is performed by rotating and translating
the pre-mirror and by rotating the grating. The translation of the
mirror implies that the optical path is wavelength dependent.
Evidently, the scanning will result on a wavelength dependent
optical path. The tunability of the path length in the magnetic
chicane  will be required to compensate for the change in the
optical path in the range of  $0.1$ mm. The last optical component
is a planar VLS grating. A monochromator resolving power of about
$4500$ (at $1.5$ nm)  is obtained with the grating having a line
spacing of $0.8$ micron, a groove height of $11$ nm and using a $3$
microns exit slit \cite{FENG,FENG2}. The estimated grating
efficiency is about $10 \%$. Reflectivity of the grazing incident
mirrors and grating are close to $100 \%$.

\section{FEL simulations}

With reference to Fig. \ref{soft1}, we performed a feasibility study
with the help of the FEL code GENESIS 1.3 \cite{GENE} running on a
parallel machine. We will present a feasibility study for the SASE3
FEL line of the European XFEL, based on a statistical analysis
consisting of $100$ runs. The overall beam parameters used in the
simulations are presented in Table \ref{tt1}.

\begin{table}
\caption{Parameters for the mode of operation at the European XFEL
used in this paper.}

\begin{small}\begin{tabular}{ l c c}
\hline & ~ Units &  ~ \\ \hline
Undulator period      & mm                  & 65     \\
Periods per cell      & -                   & 77   \\
K parameter (rms)     & -                   & 4.2  \\
Total number of cells & -                   & 21    \\
Intersection length   & m                   & 1.1   \\
Wavelength            & nm                  & 1.5  \\
Energy                & GeV                 & 10.0 \\
Charge                & nC                  & 0.1\\
\hline
\end{tabular}\end{small}
\label{tt1}
\end{table}

\begin{figure}[tb]
\includegraphics[width=0.5\textwidth]{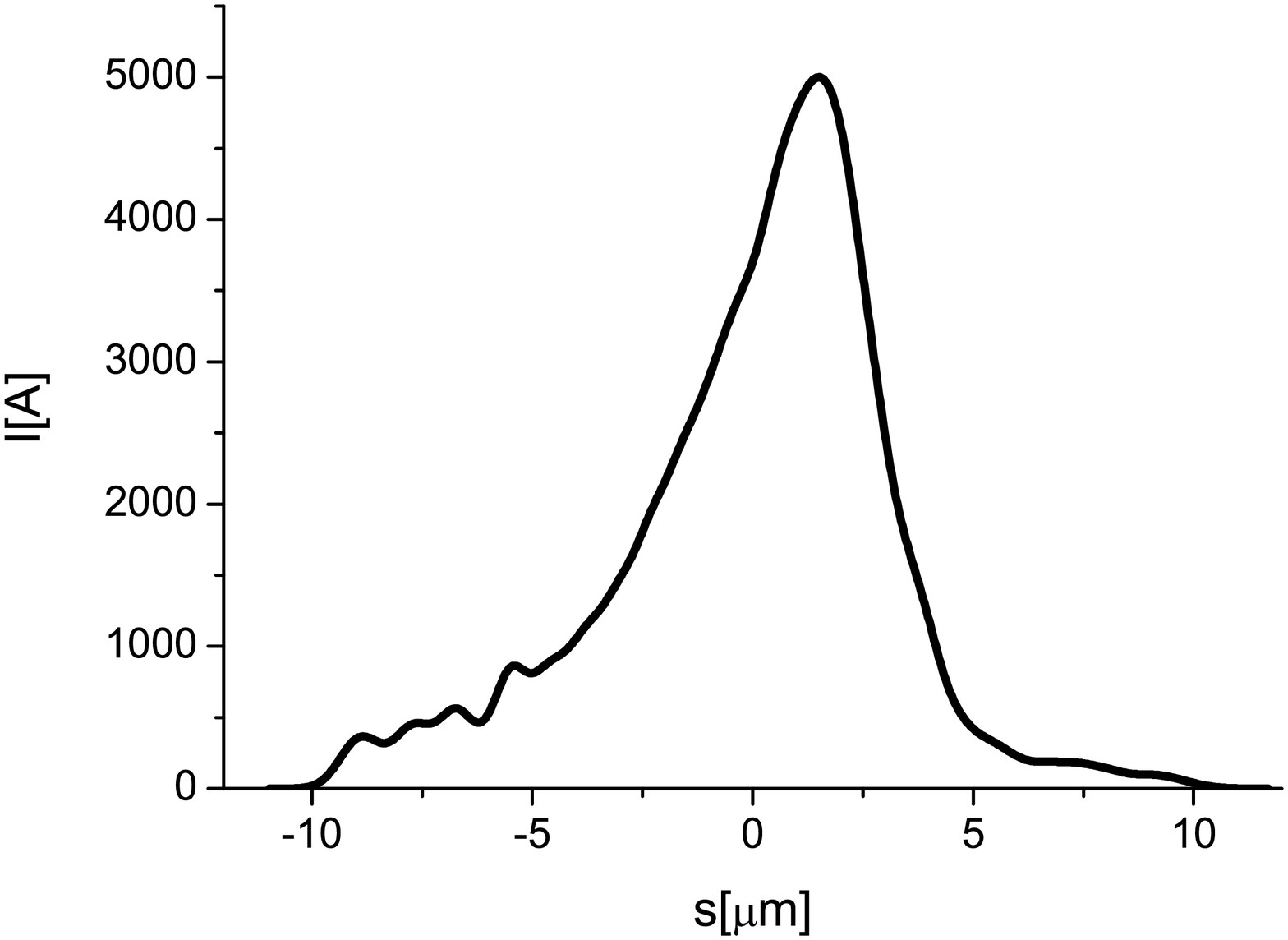}
\includegraphics[width=0.5\textwidth]{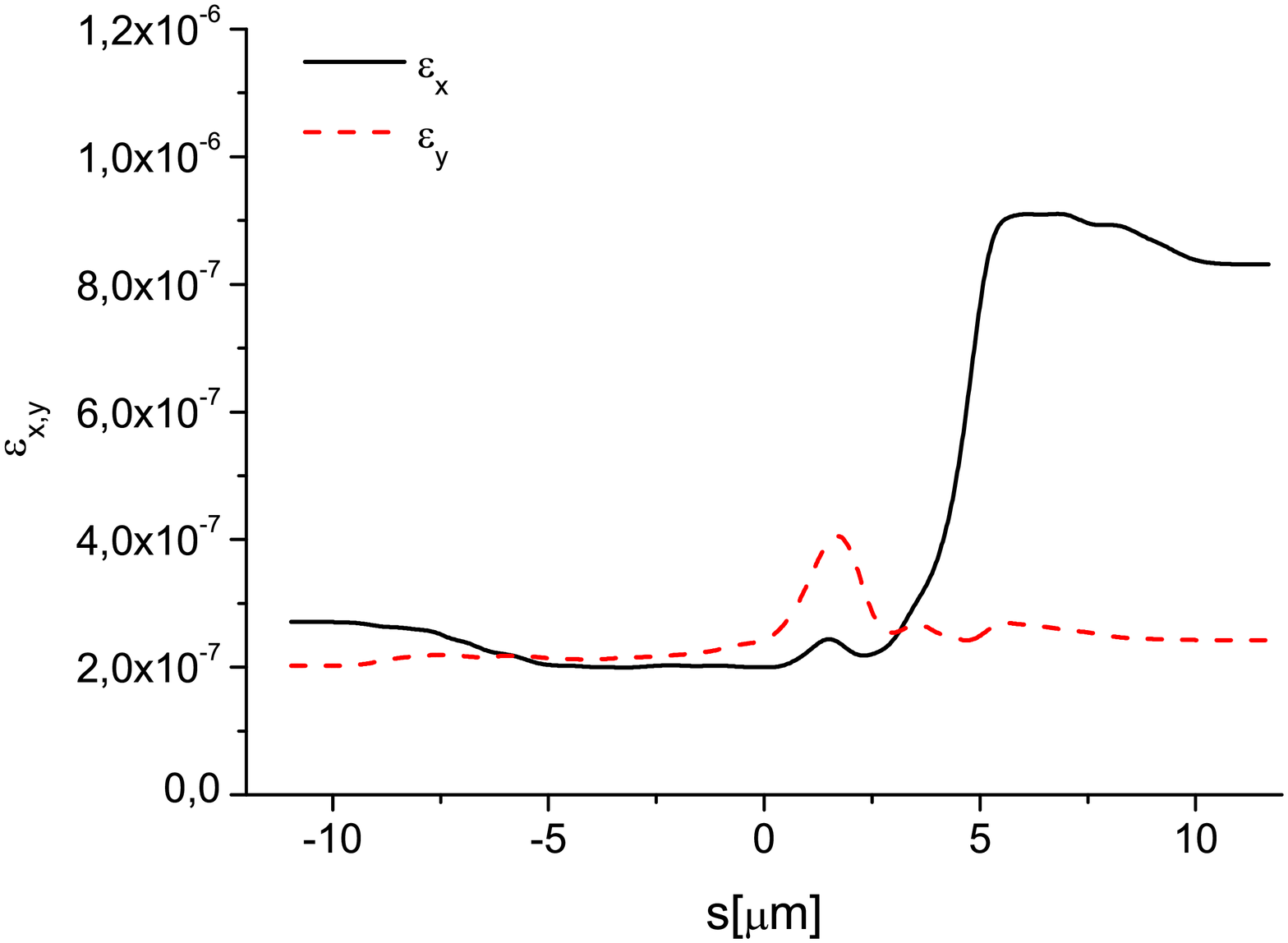}
\includegraphics[width=0.5\textwidth]{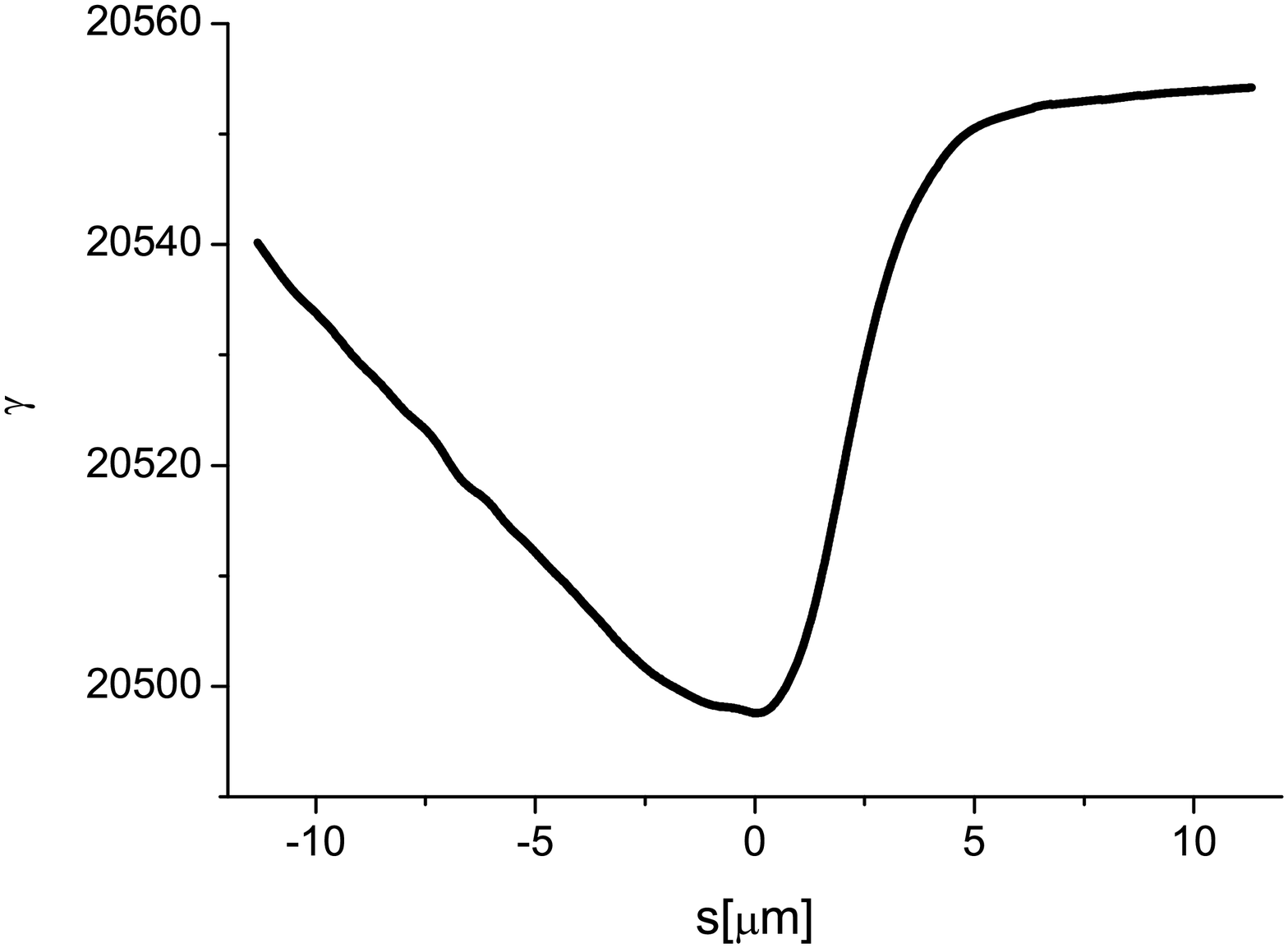}
\includegraphics[width=0.5\textwidth]{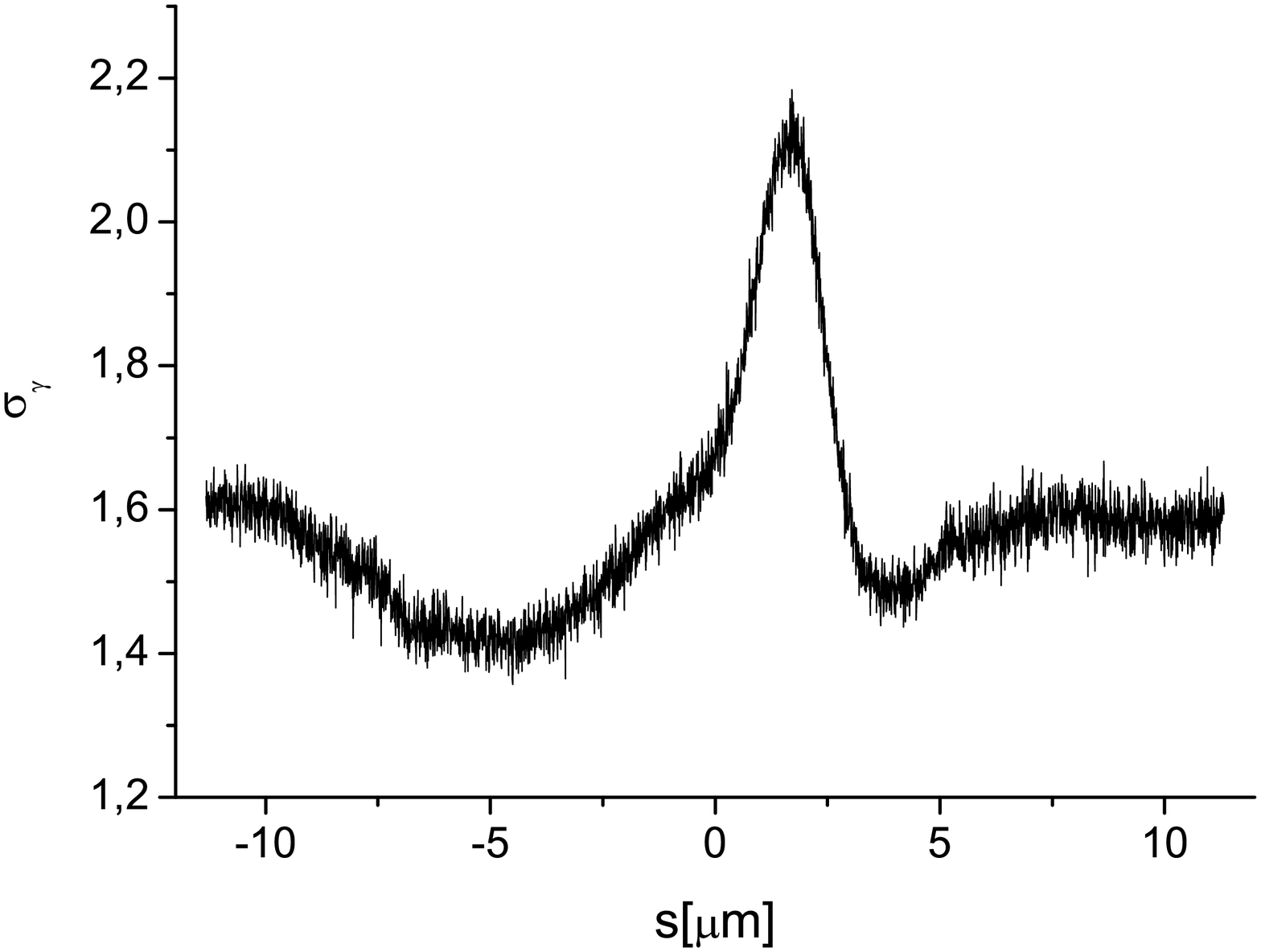}
\begin{center}
\includegraphics[width=0.5\textwidth]{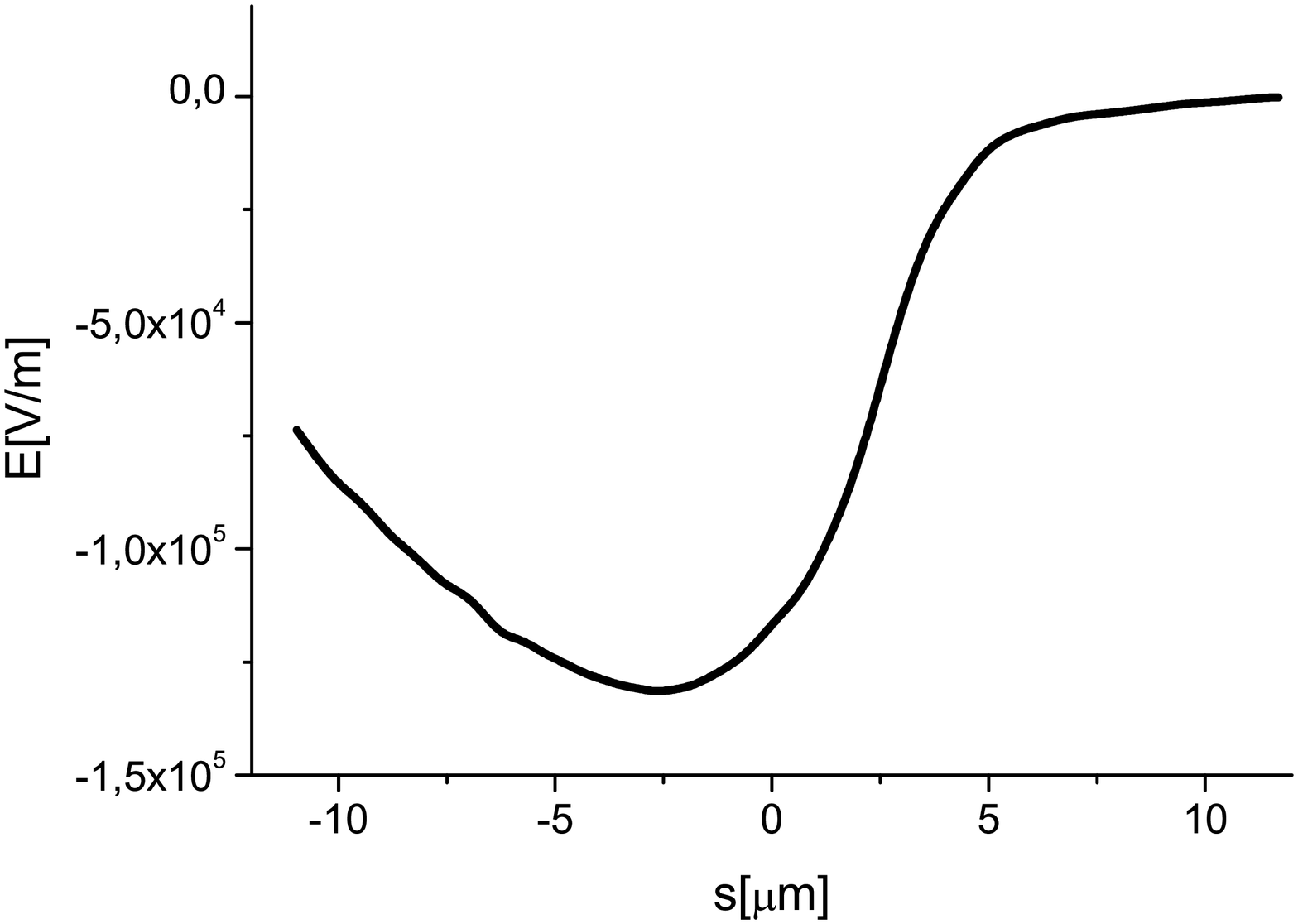}
\end{center}
\caption{Results from electron beam start-to-end simulations at the
entrance of SASE3 \cite{S2ER}. (First Row, Left) Current profile.
(First Row, Right) Normalized emittance as a function of the
position inside the electron beam. (Second Row, Left) Energy profile
along the beam. (Second Row, Right) Electron beam energy spread
profile. (Bottom row) Resistive wakefields in the SASE3 undulator
\cite{S2ER}.} \label{s2E}
\end{figure}
\begin{figure}[tb]
\includegraphics[width=0.5\textwidth]{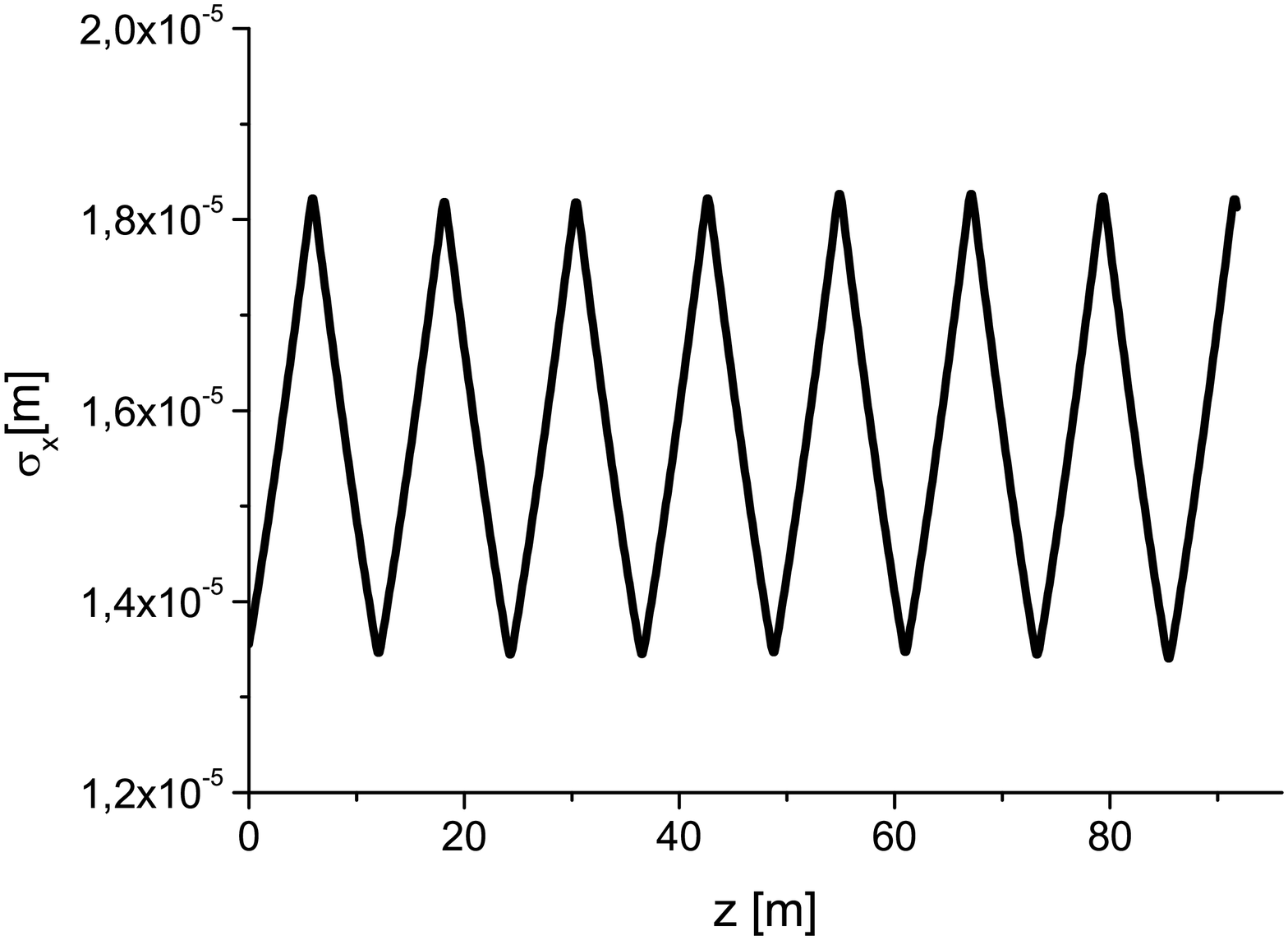}
\includegraphics[width=0.5\textwidth]{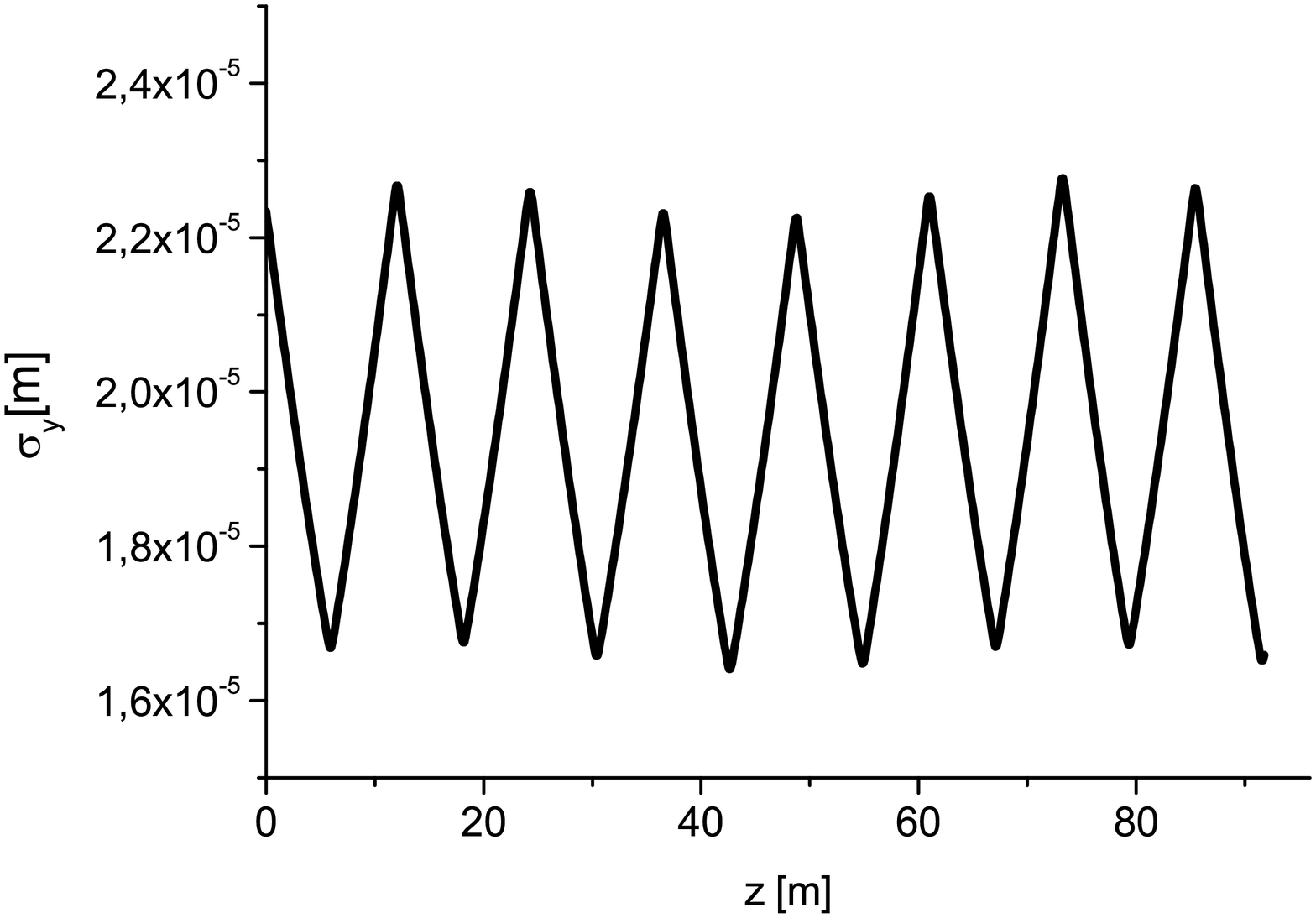}
\caption{Evolution of the horizontal (left plot) and vertical (right
plot) dimensions of the electron bunch as a function of the distance
inside the SASE3 undulator. The plots refer to the longitudinal
position inside the bunch corresponding to the maximum current
vale.} \label{sigma}
\end{figure}
The expected beam parameters at the entrance of the SASE3 undulator,
and the resistive wake inside the undulator are shown in Fig.
\ref{s2E}, \cite{S2ER}. The evolution of the transverse electron
bunch dimensions are plotted in Fig. \ref{sigma}.

\begin{figure}[tb]
\includegraphics[width=0.5\textwidth]{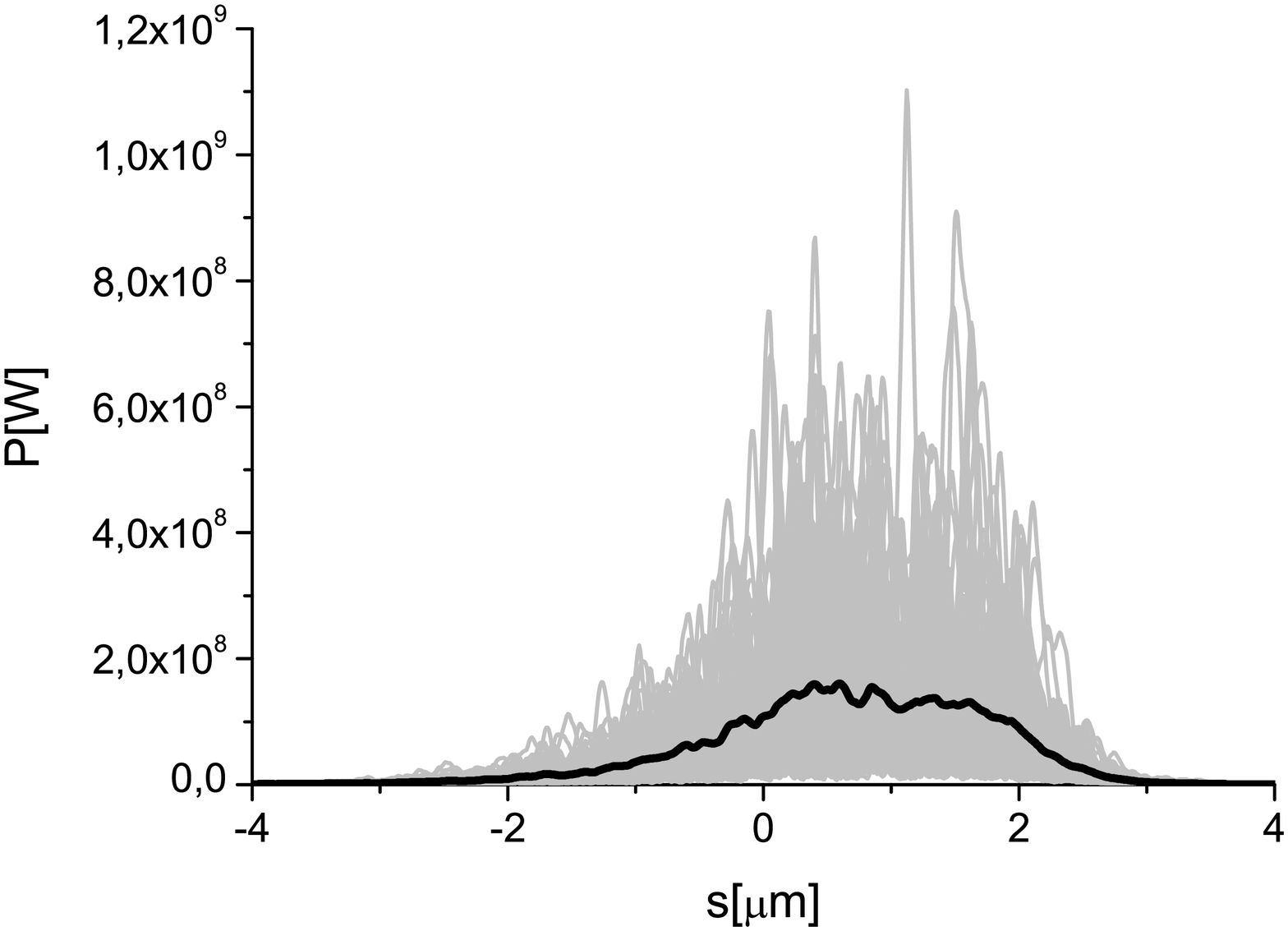}
\includegraphics[width=0.5\textwidth]{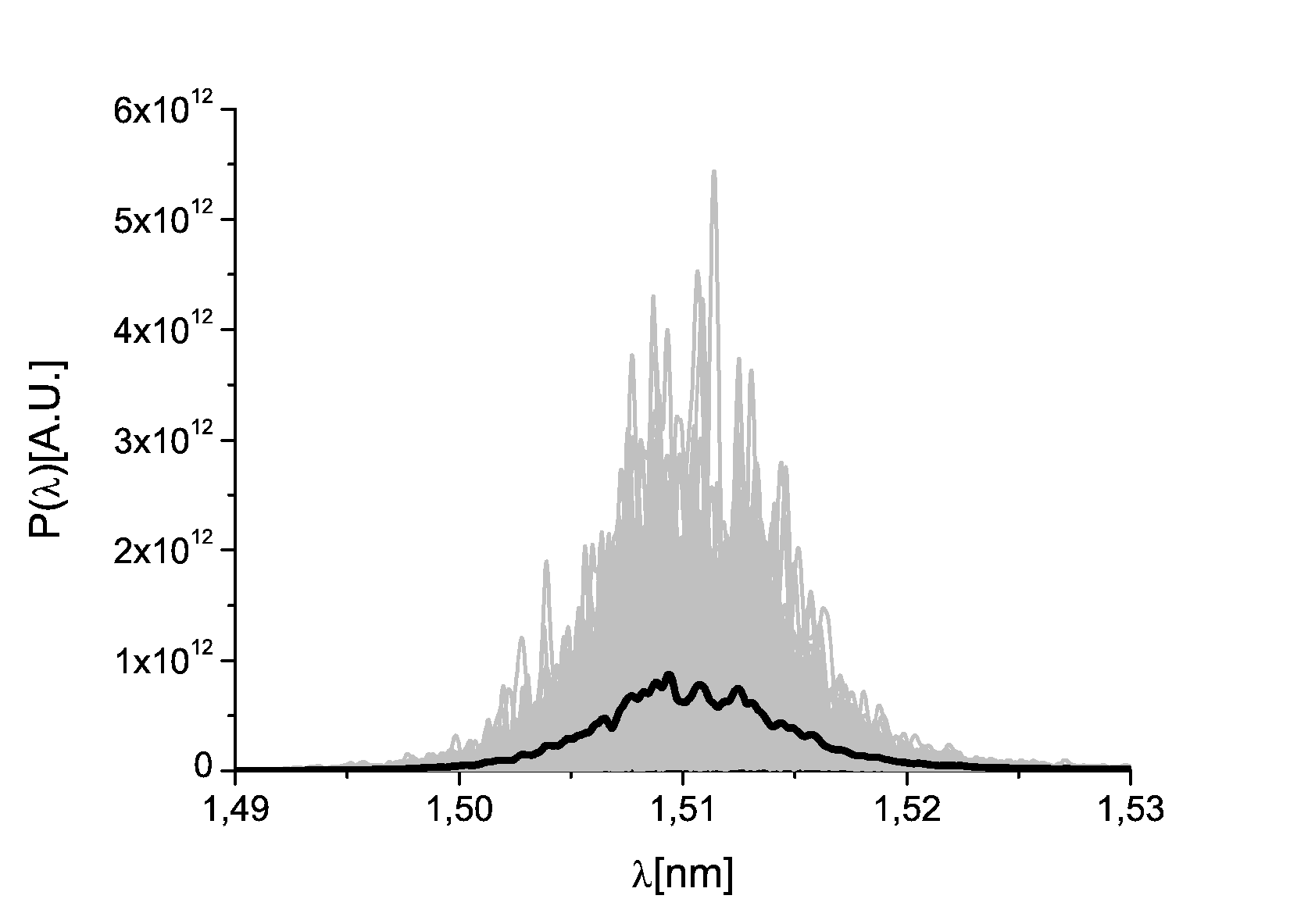}
\caption{Power distribution  and spectrum of the X-ray radiation
pulse after the first undulator. Grey lines refer to single shot
realizations, the black line refers to the average over a hundred
realizations.} \label{IN1}
\end{figure}
\begin{figure}[tb]
\includegraphics[width=0.5\textwidth]{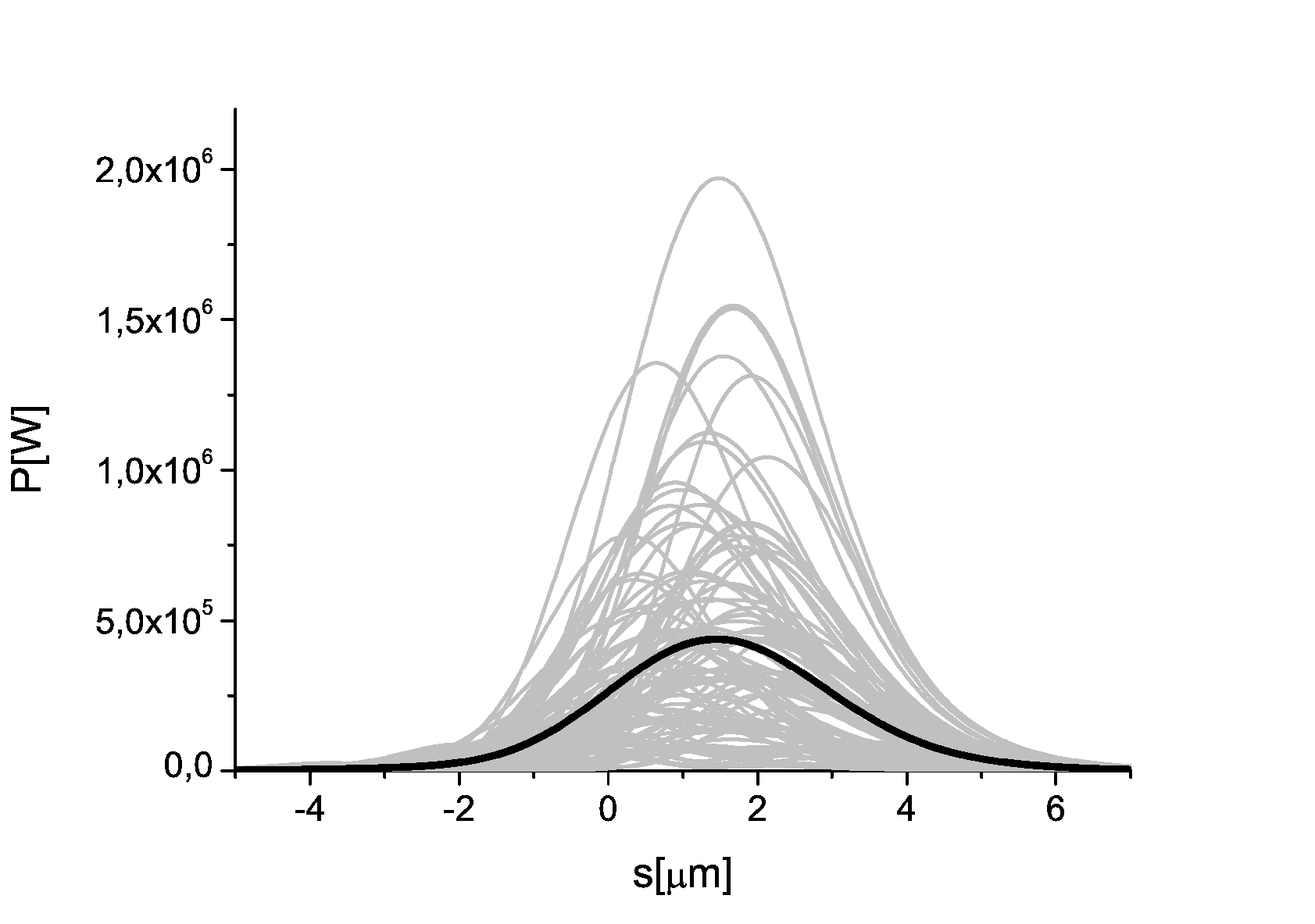}
\includegraphics[width=0.5\textwidth]{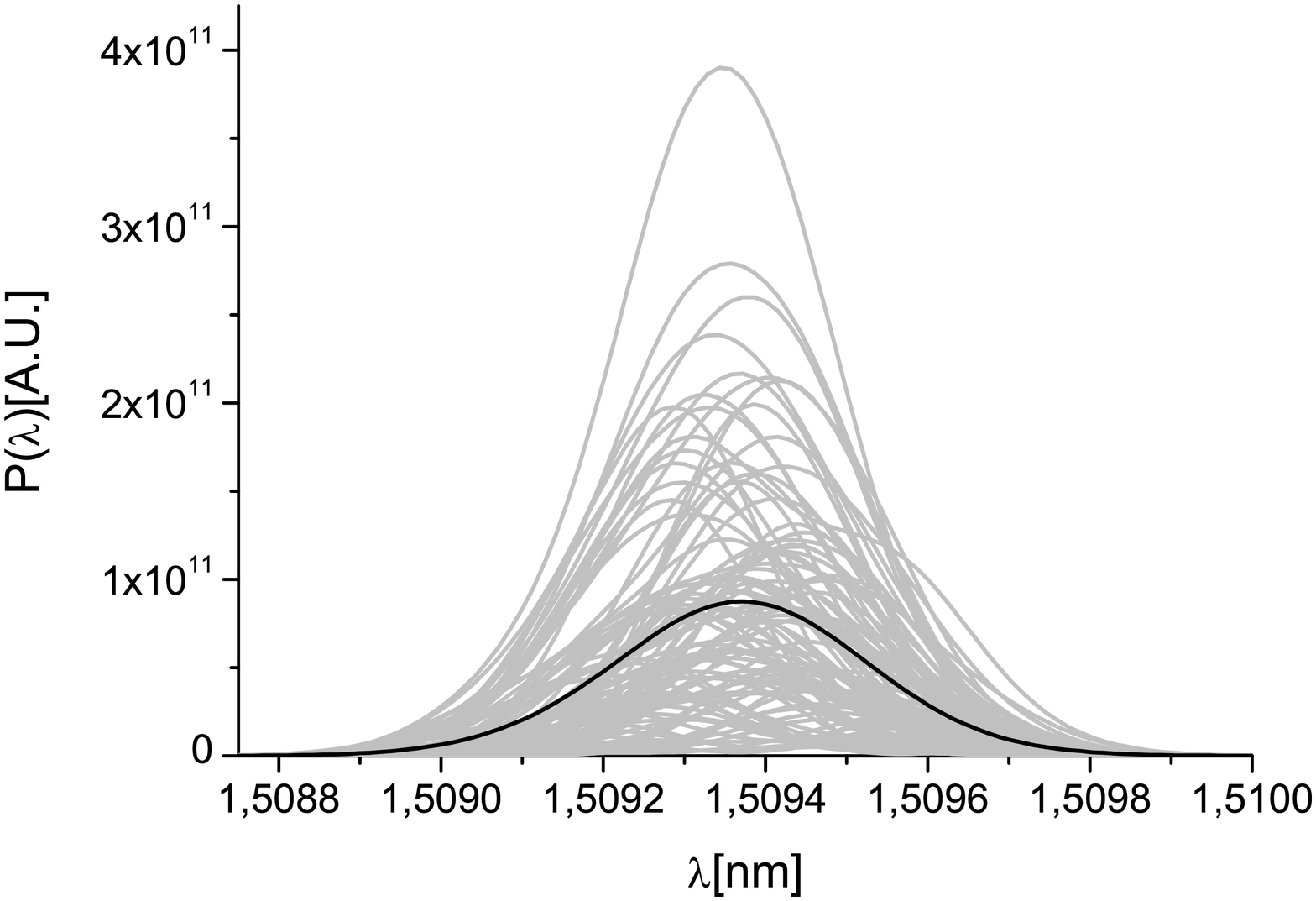}
\caption{Power distribution  and spectrum of the X-ray radiation
pulse after the monochromator. This pulse is used to seed the
electron bunch at the entrance of the output undulator. Grey lines
refer to single shot realizations, the black line refers to the
average over a hundred realizations.} \label{SEED}
\end{figure}
The SASE pulse power and spectrum after the first undulator in Fig.
\ref{soft1} is shown in Fig. \ref{IN1}. This pulse goes through the
grating monochromator. We assume a monochromator with a Gaussian
shape. At the exit of the monochromator, one obtains the seed pulse,
Fig. \ref{SEED}. As explained before, the monochromator introduces
only a short optical delay of about $2.5$ ps, which can be easily
compensated by the electron chicane. The chicane also washes out the
electron beam microbunching. As a result, at the entrance of the
second (output) undulator the electron beam and the radiation pulse
can be recombined.

\begin{figure}[tb]
\includegraphics[width=0.5\textwidth]{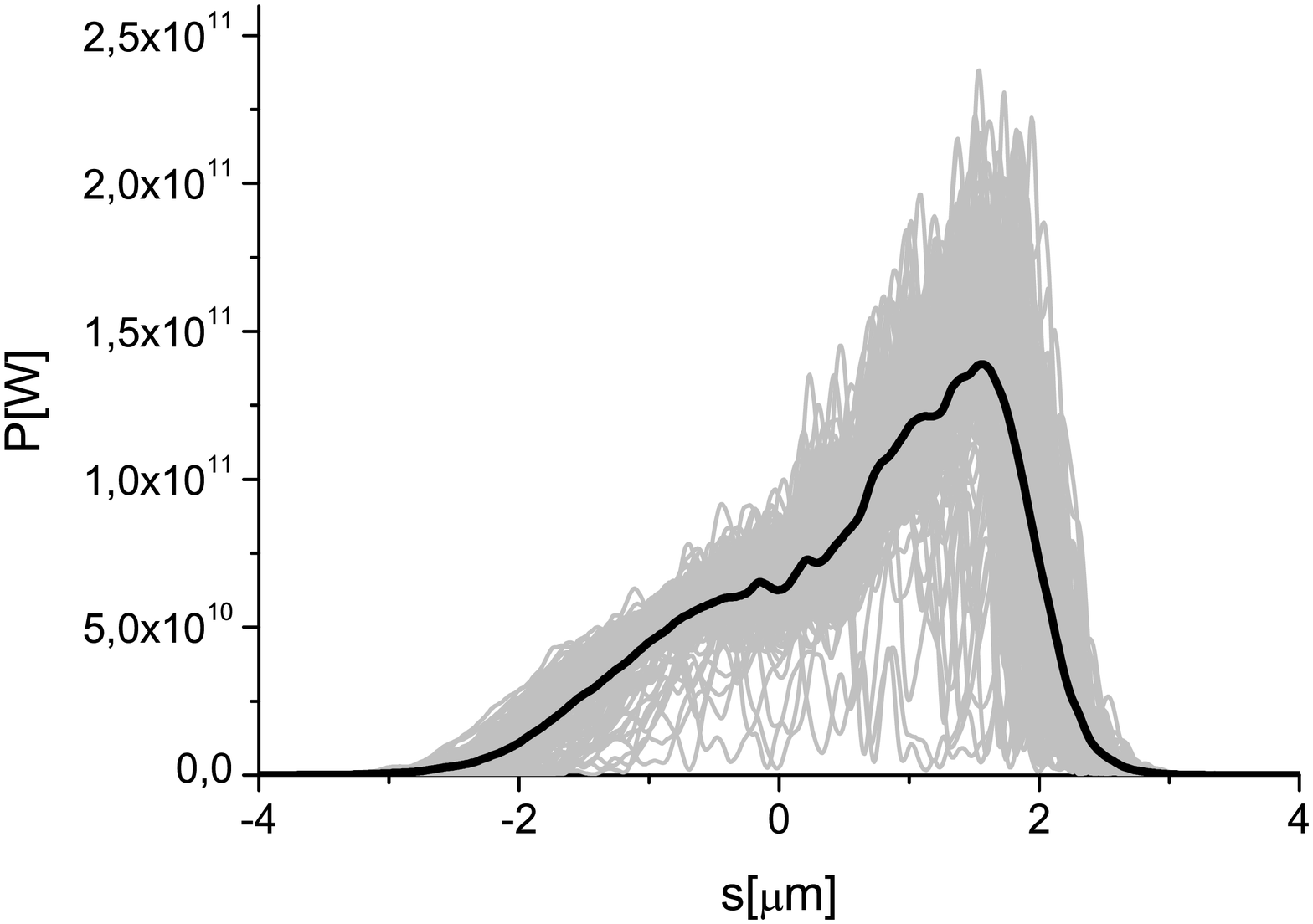}
\includegraphics[width=0.5\textwidth]{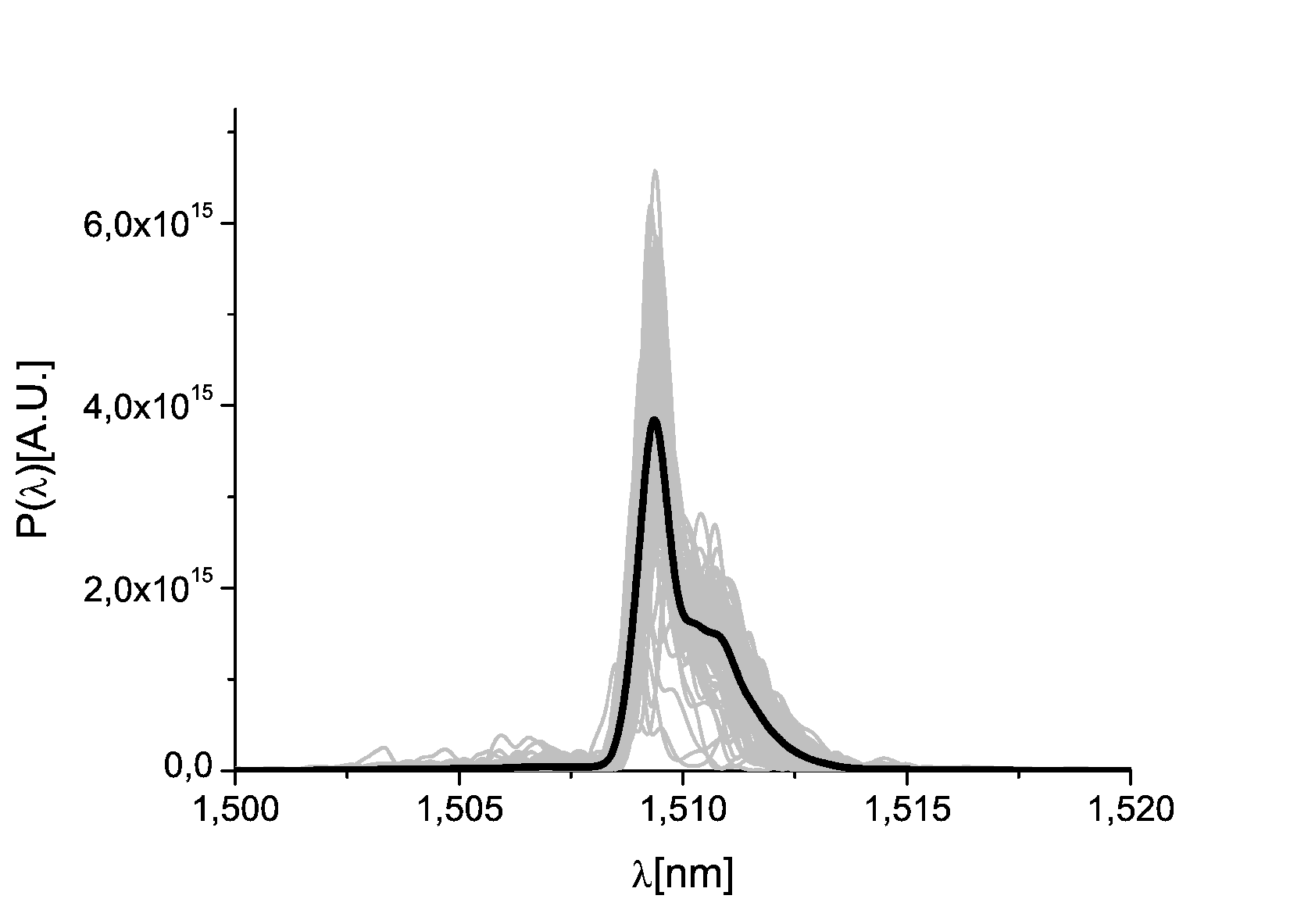}
\caption{Power distribution and spectrum of the X-ray radiation
pulse after the second undulator in the untapered case. Grey lines
refer to single shot realizations, the black line refers to the
average over a hundred realizations.} \label{OUT1}
\end{figure}
\begin{figure}[tb]
\includegraphics[width=0.5\textwidth]{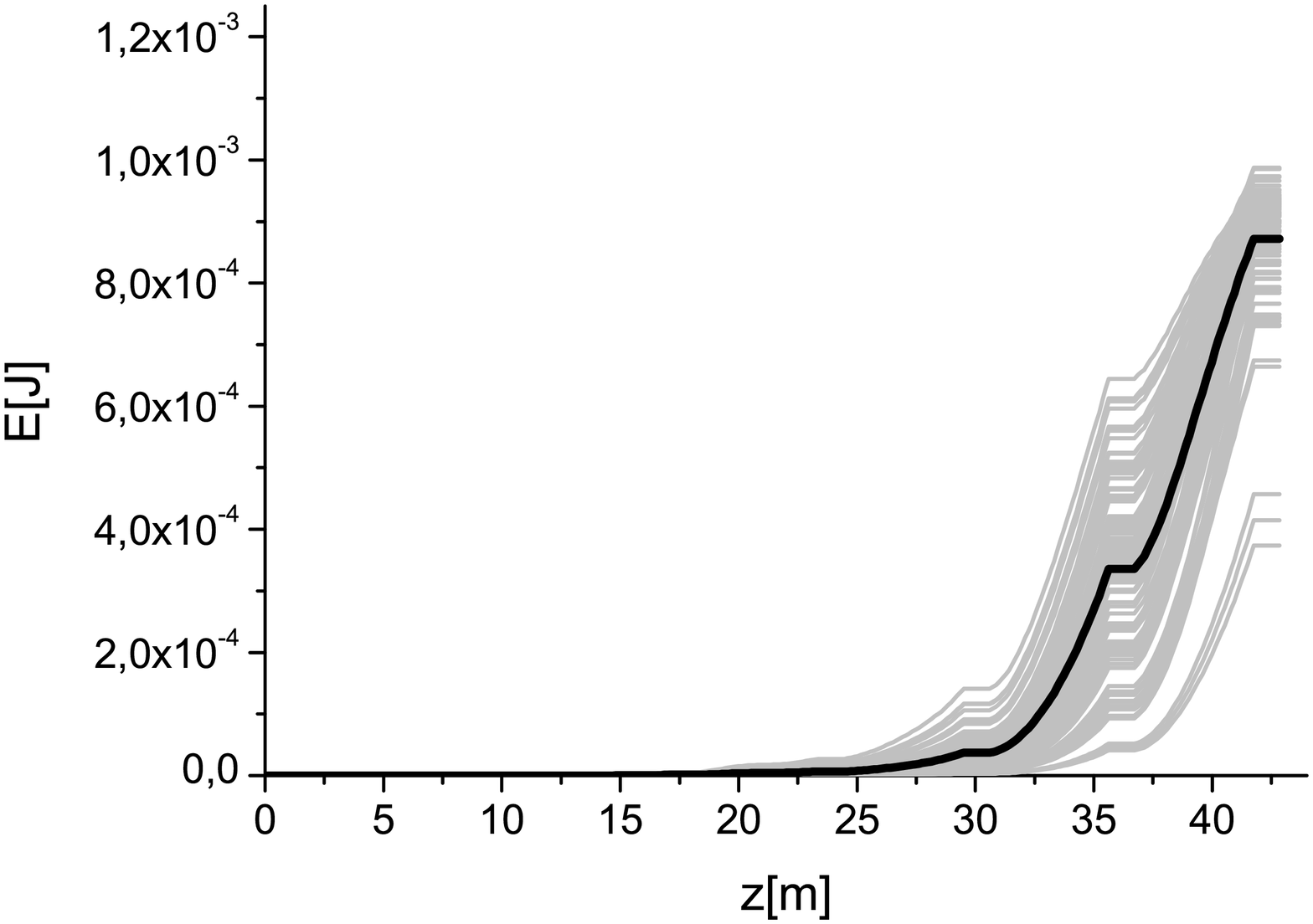}
\includegraphics[width=0.5\textwidth]{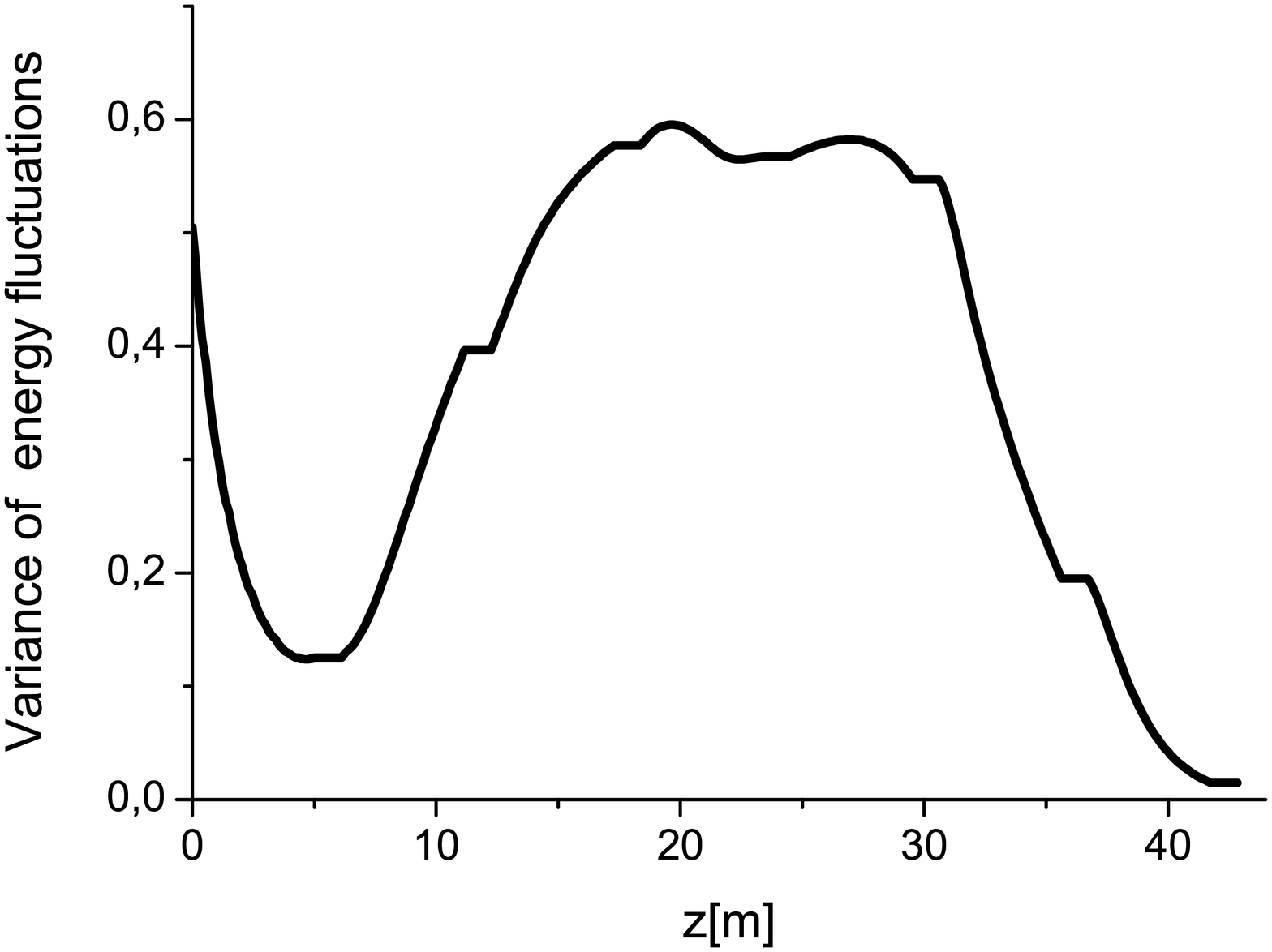}
\caption{Evolution of the energy per pulse and of the energy
fluctuations as a function of the undulator length in the untapered
case. Grey lines refer to single shot realizations, the black line
refers to the average over a hundred realizations.} \label{OUT2}
\end{figure}
\begin{figure}[tb]
\includegraphics[width=0.5\textwidth]{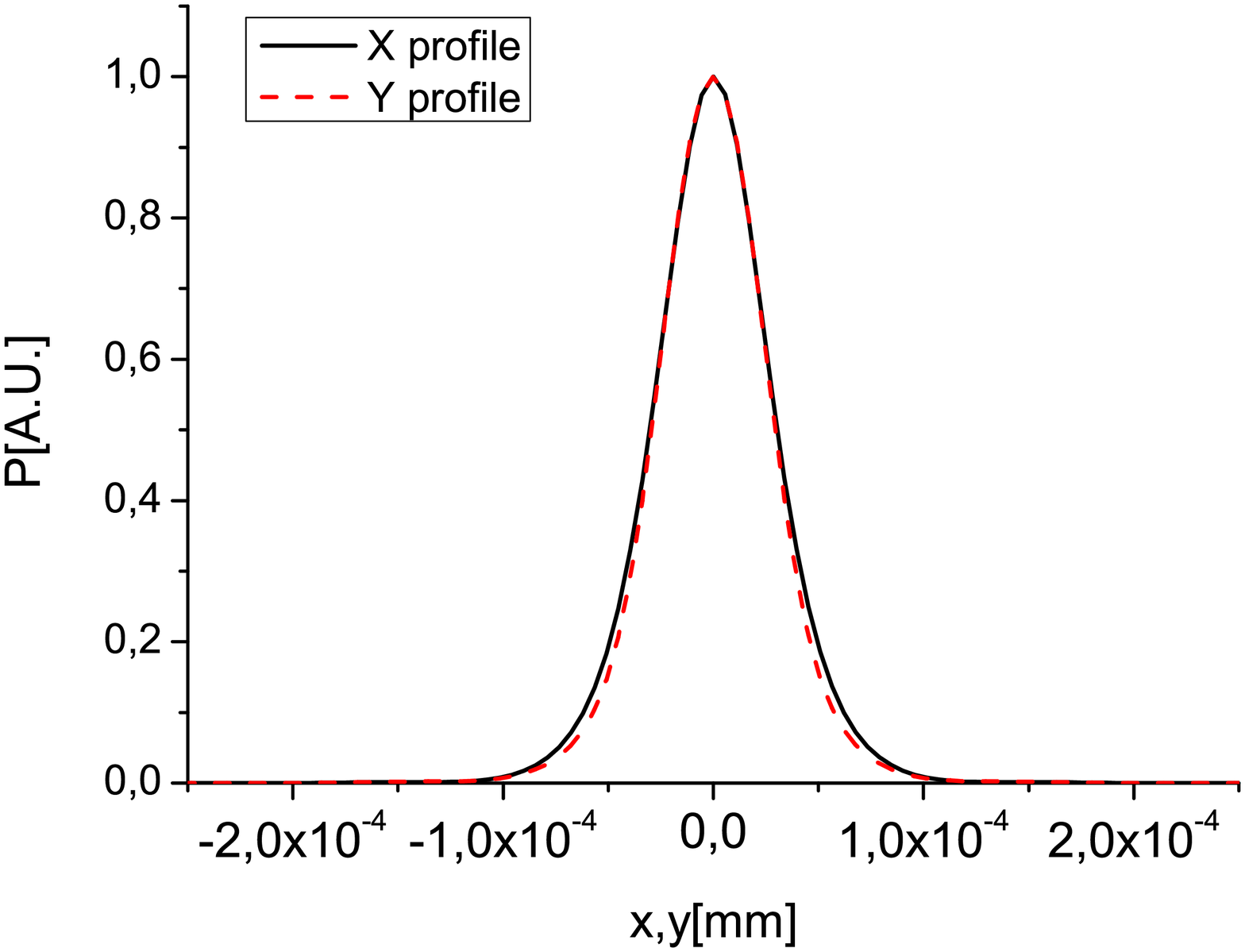}
\includegraphics[width=0.5\textwidth]{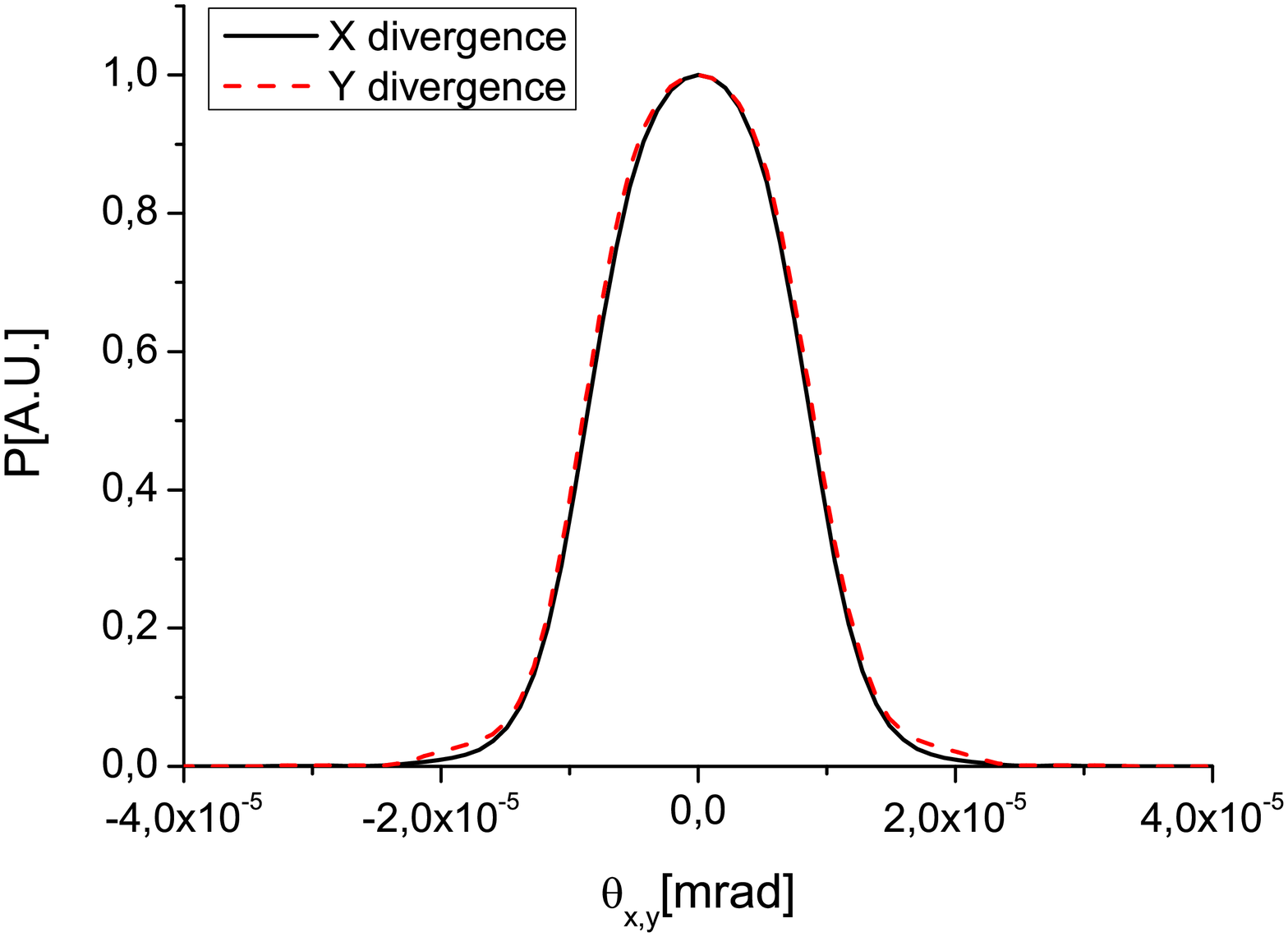}
\caption{(Left plot) Transverse radiation distribution in the
untapered case at the exit of the output undulator. (Right plot)
Directivity diagram of the radiation distribution in the case of
tapering at the exit of the output undulator.} \label{spot}
\end{figure}

If the output undulator is not tapered, one needs $7$ sections to
reach saturation. The best compromise between power and spectral
bandwidth are reached after $6$ sections, Fig. \ref{OUT1}. In this
case, the evolution of the energy per pulse and of the energy
fluctuations as a function of the undulator length are shown in Fig.
\ref{OUT2}. The pulse now reaches the $100$ GW power level, with an
average relative FWHM spectral width narrower than $10^{-3}$.
Finally, the transverse radiation distribution and divergence at the
exit of the output undulator are shown in Fig. \ref{spot}.

\begin{figure}[tb]
\begin{center}
\includegraphics[width=0.5\textwidth]{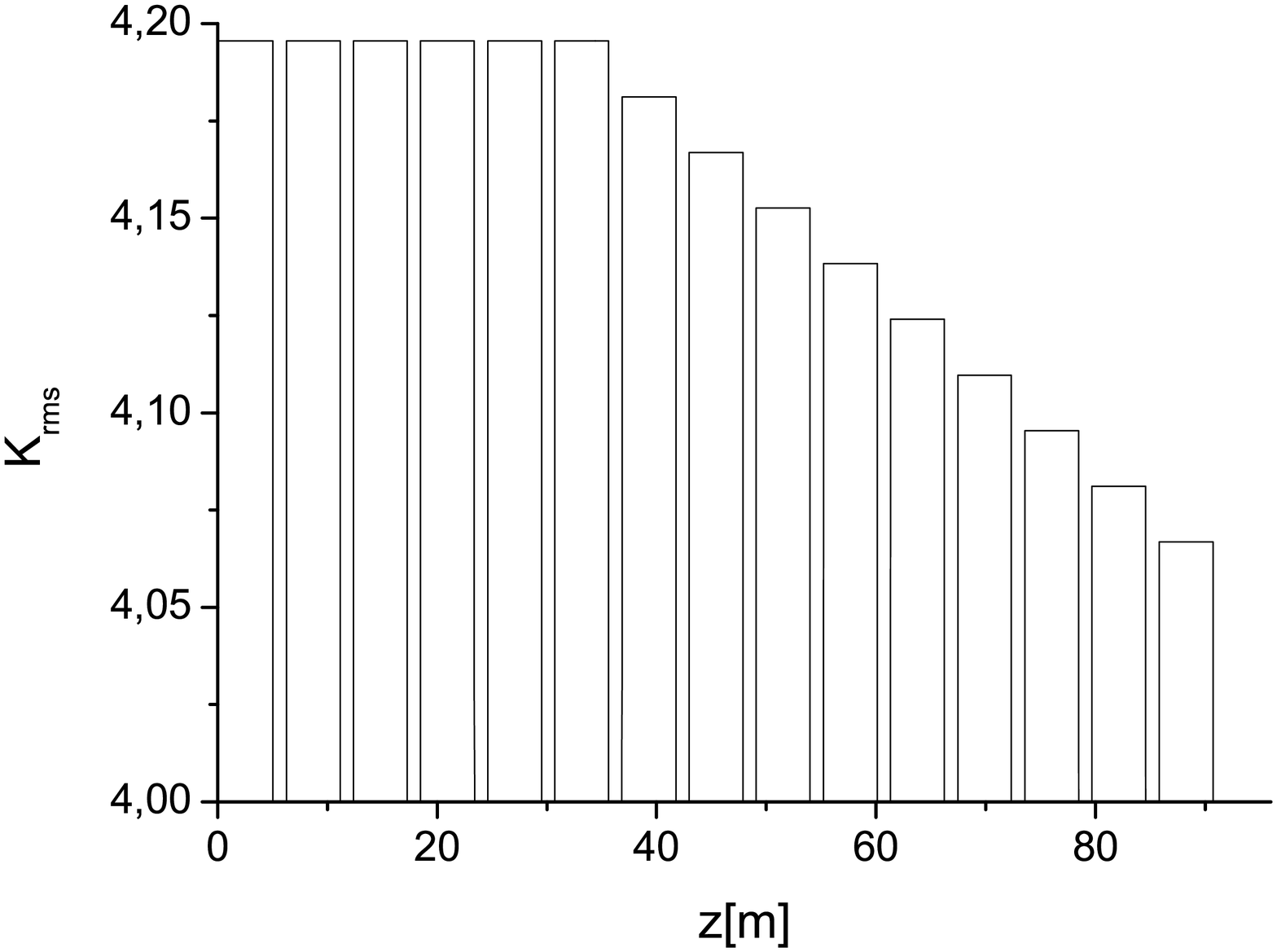}
\end{center}
\caption{Taper configuration for high-power mode of operation at
$1.5$ nm.} \label{krms}
\end{figure}
\begin{figure}[tb]
\includegraphics[width=0.5\textwidth]{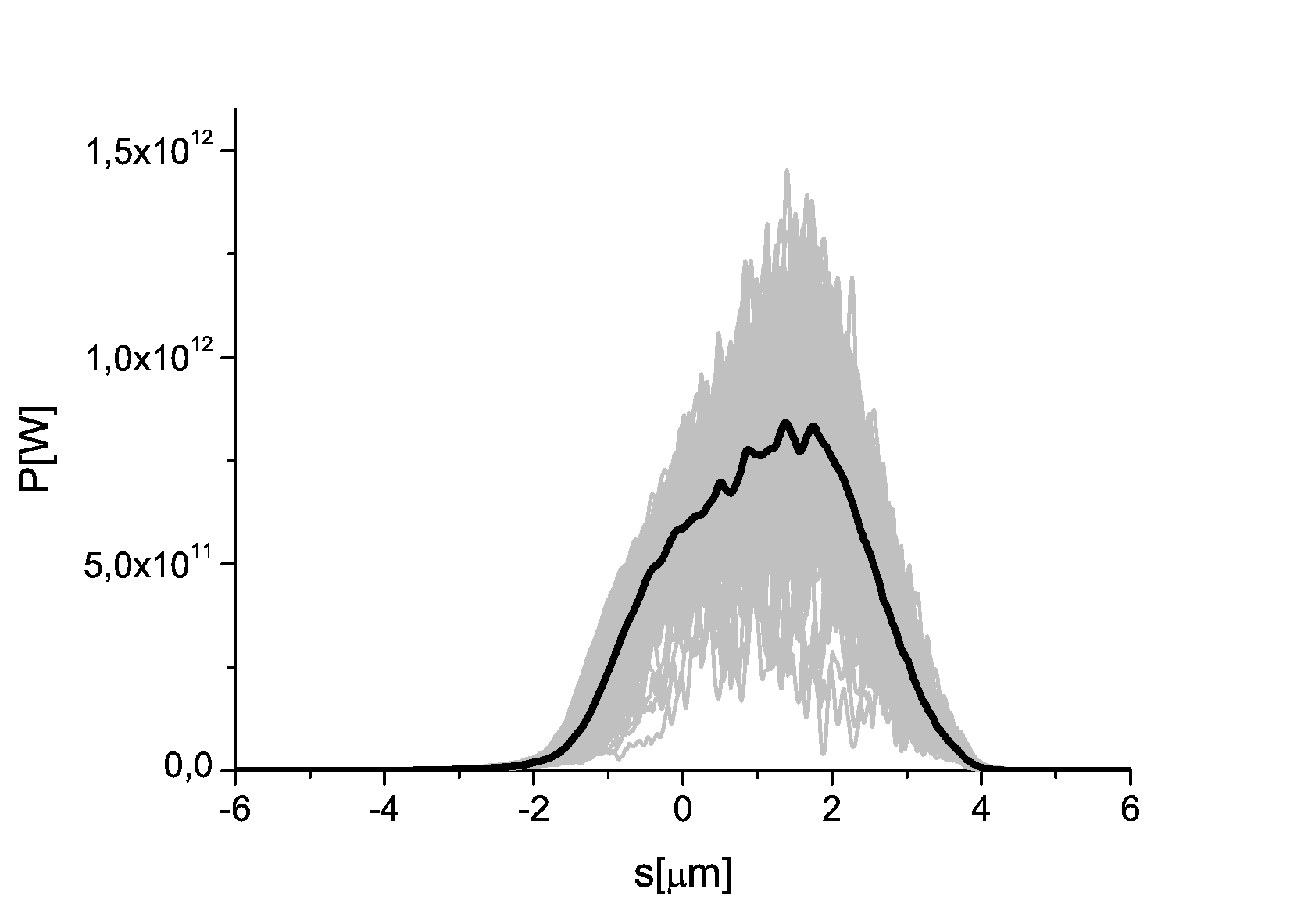}
\includegraphics[width=0.5\textwidth]{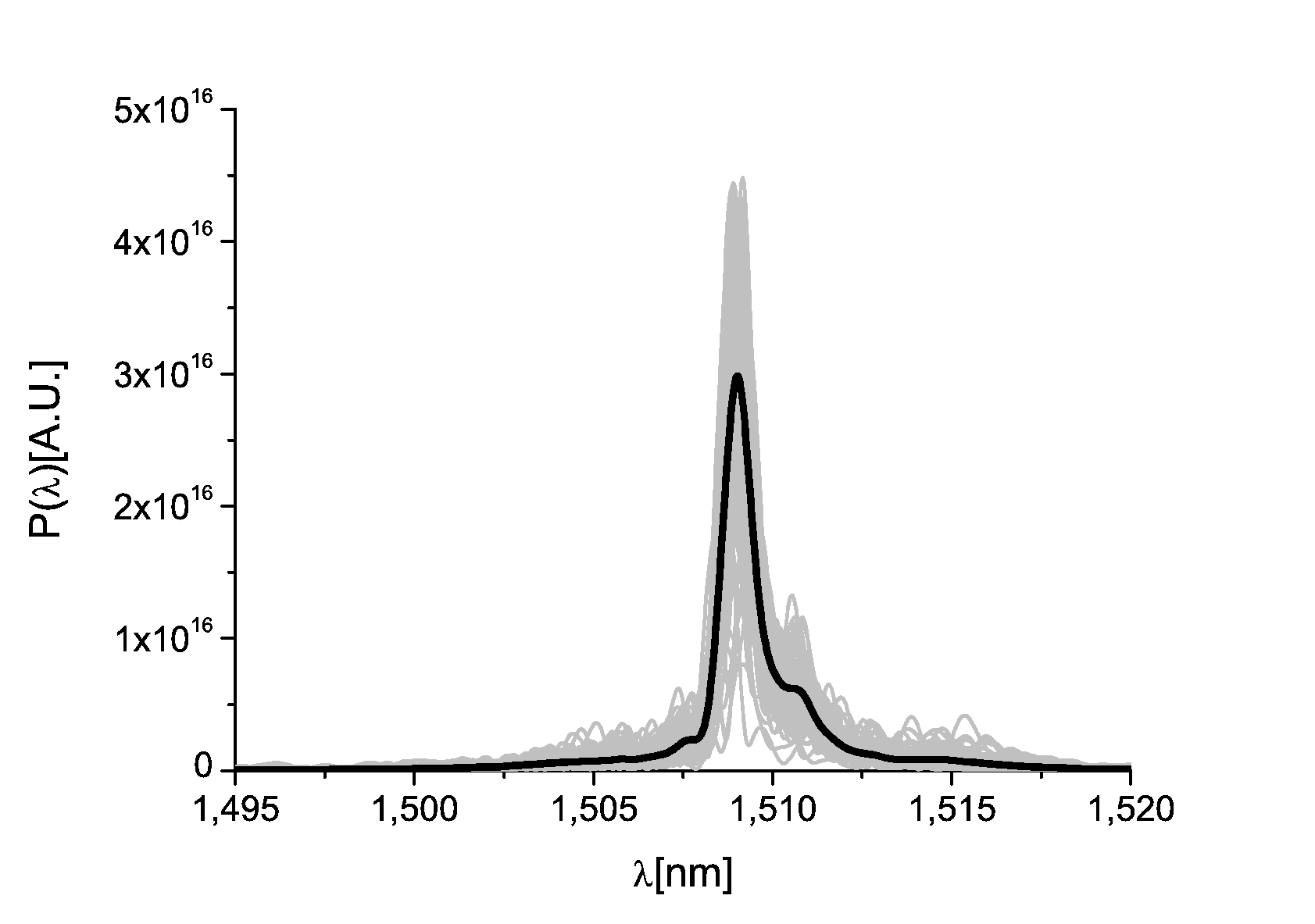}
\caption{Power distribution and spectrum of the X-ray radiation
pulse after the second undulator in the tapered case. Grey lines
refer to single shot realizations, the black line refers to the
average over a hundred realizations.} \label{OUTT1}
\end{figure}
\begin{figure}[tb]
\includegraphics[width=0.5\textwidth]{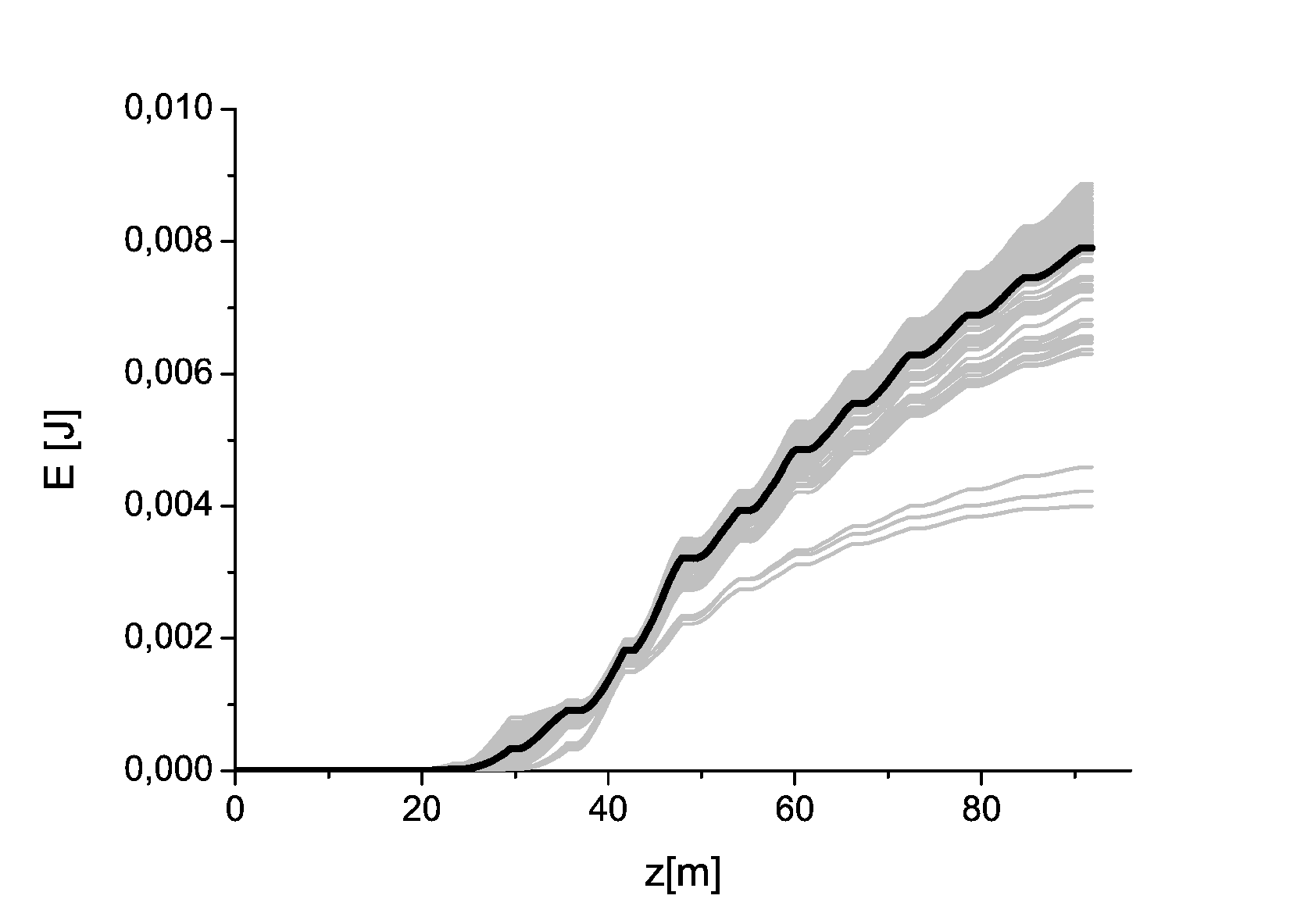}
\includegraphics[width=0.5\textwidth]{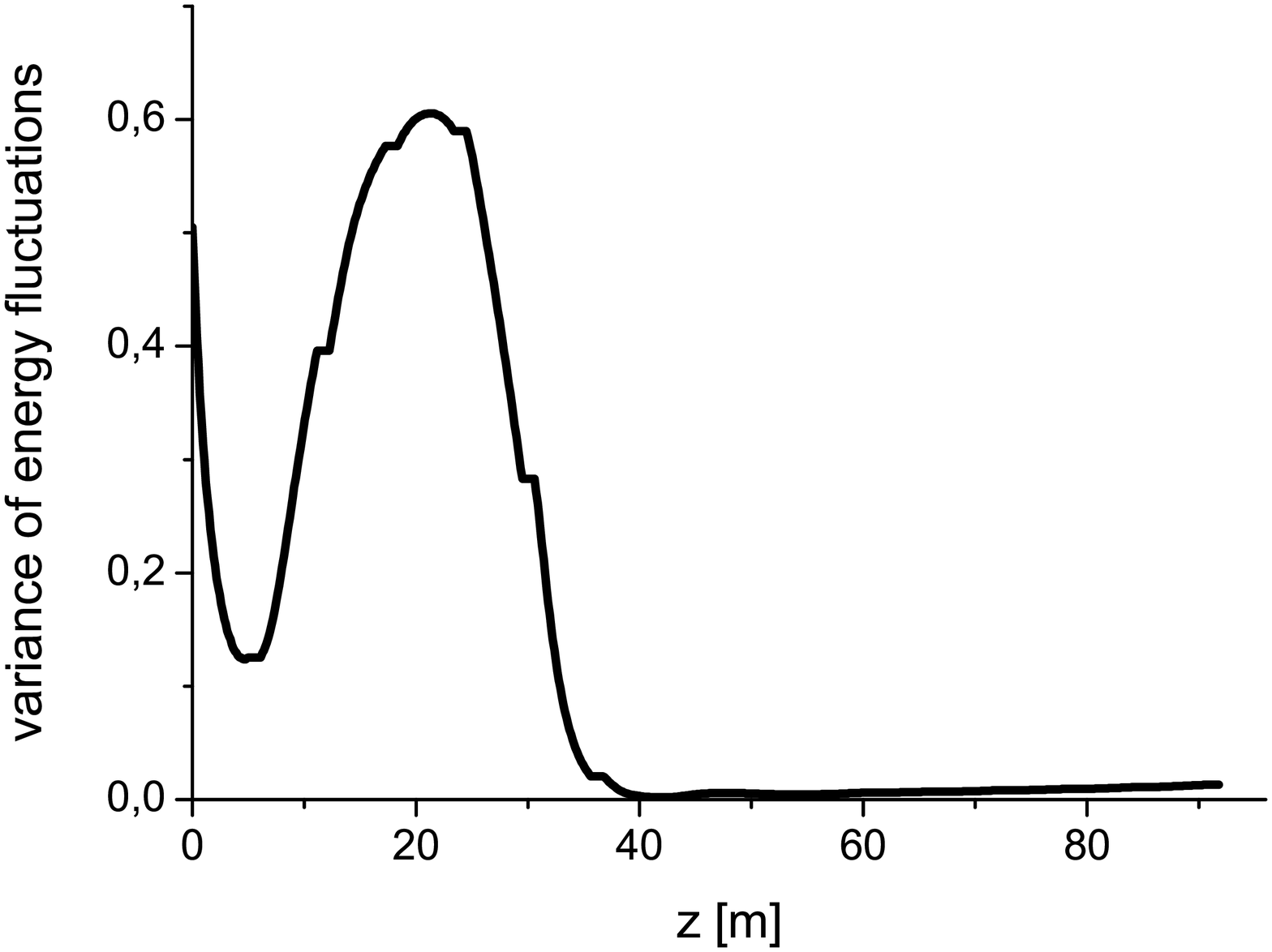}
\caption{Evolution of the energy per pulse and of the energy
fluctuations as a function of the undulator length in the tapered
case. Grey lines refer to single shot realizations, the black line
refers to the average over a hundred realizations.} \label{OUTT2}
\end{figure}
\begin{figure}[tb]
\includegraphics[width=0.5\textwidth]{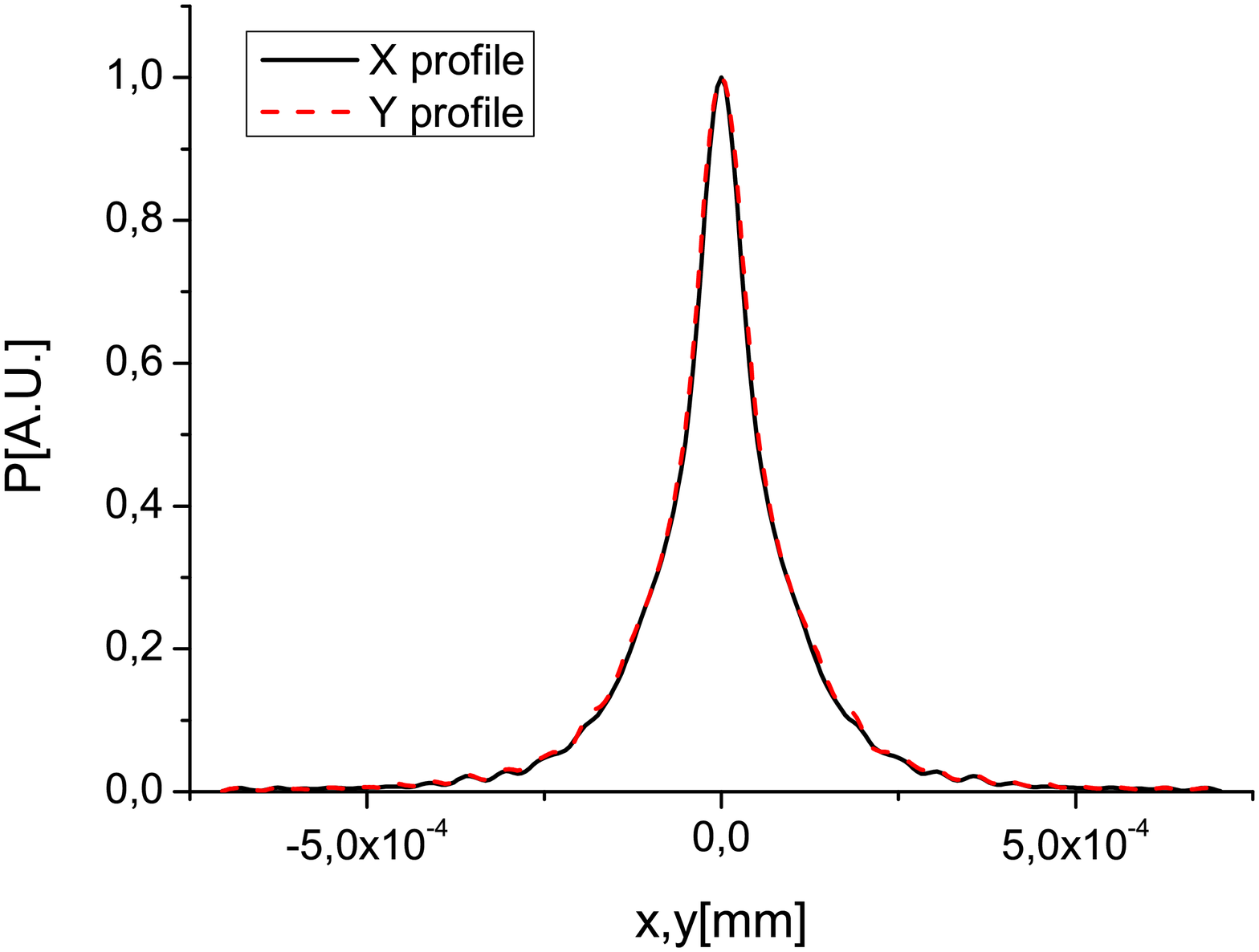}
\includegraphics[width=0.5\textwidth]{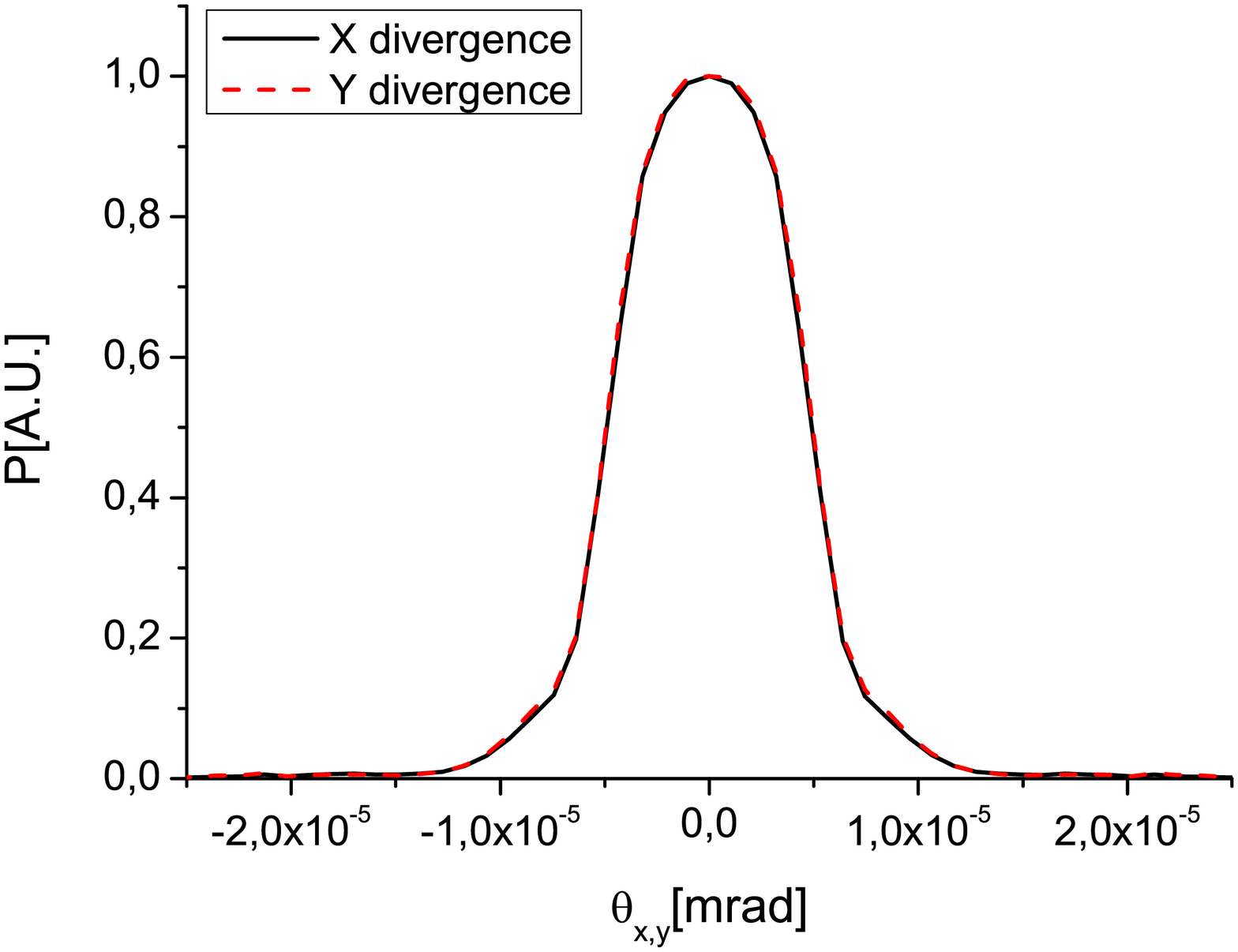}
\caption{(Left plot) Transverse radiation distribution in the case
of tapering at the exit of the output undulator. (Right plot)
Directivity diagram of the radiation distribution in the case of
tapering at the exit of the output undulator.} \label{spotT}
\end{figure}
The most promising way to increase the output power is via
post-saturation tapering. Tapering consists in a slow reduction of
the field strength of the undulator in order to preserve the
resonance wavelength, while the kinetic energy of the electrons
decreases due to the FEL process. The undulator taper could be
simply implemented as a step taper from one undulator segment to the
next, as shown in Fig. \ref{krms}. The magnetic field tapering is
provided by changing the undulator gap. A further increase in power
is achievable by starting the FEL process from the monochromatic
seed, rather than from noise. The reason is the higher degree of
coherence of the radiation in the seed case, thus involving, with
tapering, a larger portion of the bunch in the energy-wavelength
synchronism. Using the tapering configuration in Fig. \ref{krms},
one obtains the output characteristics, in terms of power and
spectrum, shown in Fig. \ref{OUTT1}. The output power is increased
of about a factor ten, allowing one to reach about one TW. The
spectral width remains almost unvaried, with an average relative
bandwidth (FWHM) narrower than $10^{-3}$. The evolution of the
energy per pulse and of the energy fluctuations as a function of the
undulator length are shown in Fig. \ref{OUT2}. Finally, the
transverse radiation distribution and divergence at the exit of the
output undulator are shown in Fig. \ref{spotT}. By comparison with
Fig. \ref{spot} one can see that the divergence decrease is
accompanied by an increase in the transverse size of the radiation
spot at the exit of the undulator.

\section{Conclusions}

In this paper we showed that monochromatization of soft X-ray pulses
is of great importance for the European XFEL upgrade program. In
fact, aside for an improvement of the longitudinal coherence, highly
monochromatized pulses open up many possibilities to enhance the
capacity of the European XFEL.

In particular we showed that monochromatization effectively allows
one to use undulator tapering techniques enabling a TW-power mode of
operation in the soft X-ray wavelength regime.

The progress which can be achieved with the methods considered here
is based on the novel soft X-ray self-seeding scheme with grating
monochromator  \cite{FENG,FENG2}, which is extremely compact and can
be straightforwardly realized at the European XFEL baseline
undulator SASE3. Preliminary estimates seem to indicate that a
monochromator efficiency close to $10 \%$ can be achieved. However,
detailed efficiency calculations are required to determine if this
is indeed the case. Another problem with the operation of this
scheme can be the mismatching of the seed radiation pulse with the
electron bunch. However, the safety margin of the proposed design is
large enough: even if the nominal seed power (200 kW) is degraded of
an order of magnitude due to non-ideal effects, one would still
obtain a reasonable seed signal (20 kW) compared to the equivalent
shot noise power (1 kW).

\section{Acknowledgements}

We are grateful to Massimo Altarelli, Reinhard Brinkmann,
Serguei Molodtsov and Edgar Weckert for their support and their interest during the compilation of this work.


\begin{thebibliography}{99}

\bibitem{FENG} Y. Feng, J. Hastings, P. Heimann, M. Rowen, J. Krzywinski, and J. Wu, "X-ray Optics for soft X-ray self-seeding
the LCLS-II", proceedings of 2010 FEL conference, Malmo, Sweden,
(2010).

\bibitem{FENG2} Y. Feng, P. Heimann, J. Wu, J. Krzywinski, M. Rowen, and J. Hastings,
"Compact Grating Monochromator Design for LCLS-I Soft X-ray
Self-Seeding", https://slacportal.slac.stanford.edu/
sites/lcls$\_$public/lcls$\_$ii/Lists/LCLS$\_$II$\_$
 Calendar/Physics$\_$Meetings.aspx, May 2011 and
https://sites.google.com/a/lbl.gov/realizing-the-potential-of-seeded-fels-in-the-soft-x-ray-regime-workshop/talks,
October 2011

\bibitem{LCLS2} P. Emma et al., Nature photonics doi:10.1038/nphoton.2010.176 (2010)

\bibitem{SPRIN} T. Tanaka et al. (Eds.) Spring-8 Compact SASE Source Conceptual Design report, Kouto
(2005) (See also http://www-xfel.spring8.or.jp/SCSSCDR.pdf)

\bibitem{tdr-2006} M. Altarelli, et al. (Eds.)
XFEL, The European X-ray Free-Electron Laser, Technical Design
Report, DESY 2006-097, Hamburg (2006).

\bibitem{SELF} J. Feldhaus et al., Optics. Comm. 140, 341 (1997).

\bibitem{SXFE} E. Saldin, E. Schneidmiller,  Yu. Shvyd'ko and M.
Yurkov, NIM A 475 357 (2001).

\bibitem{SOPT} E. Saldin, E. Schneidmiller and M. Yurkov, NIM A 445
178 (2000).

\bibitem{STTF} R. Treusch, W. Brefeld, J. Feldhaus and U Hahn,  Ann. report 2001
"The seeding project for the FEL in TTF phase II" (2001).

\bibitem{SCOM} A. Marinelli et al., Comparison of HGHG and Self Seeded Scheme for the Production of Narrow Bandwidth FEL
Radiation, Proceedings of FEL 2008, MOPPH009, Gyeongju (2008).

\bibitem{OURL} G. Geloni, V. Kocharyan and E.~Saldin, "Scheme for generation of highly monochromatic X-rays from a baseline
XFEL  undulator", DESY 10-033 (2010).

\bibitem{HUAN} Y. Ding, Z. Huang and R. Ruth, Phys.Rev.ST Accel.Beams, vol. 13, p. 060703 (2010).


\bibitem{OURX} G. Geloni, V. Kocharyan and E.~Saldin, "A simple method for controlling the line width of SASE X-ray FELs",
DESY 10-053 (2010).

\bibitem{OURY2} G. Geloni, V. Kocharyan and E.~Saldin, "A Cascade self-seeding scheme with wake monochromator for narrow-bandwidth X-ray FELs", DESY 10-080 (2010).


\bibitem{OURY4} Geloni, G., Kocharyan, V., and Saldin, E., "Cost-effective way to enhance the capabilities of the LCLS baseline", DESY 10-133 (2010).


\bibitem{OURY5b} Geloni, G., Kocharyan V., and Saldin, E., "A novel Self-seeding scheme for hard X-ray FELs", Journal of Modern
Optics, DOI:10.1080/09500340.2011.586473


\bibitem{WU} J. Wu et al., "Staged self-seeding scheme for narrow
bandwidth , ultra-short X-ray harmonic generation free electron
laser at LCLS", proceedings of 2010 FEL conference, Malmo, Sweden,
(2010).



\bibitem{OURY3} G. Geloni, V. Kocharyan and E.~Saldin, "Scheme for generation of fully coherent, TW power level hard x-ray pulses from baseline undulators at the European XFEL", DESY 10-108 (2010).

\bibitem{OURY5} Geloni, G.,  Kocharyan, V.,  and Saldin, E., "Production of transform-limited X-ray pulses through
self-seeding at the European X-ray FEL", DESY 11-165 (2011).


\bibitem{WUFEL1} W.M. Fawley et al., Toward TW-level LCLS radiation
pulses, TUOA4, to appear in the FEL 2011 Conference proceedings,
Shanghai, China, 2011

\bibitem{WUFEL2} J. Wu et al., Simulation of the Hard X-ray
Self-seeding FEL at LCLS, MOPB09, to appear in the FEL 2011
Conference proceedings, Shanghai, China, 2011


\bibitem{CDRL2} The LCLS-II Conceptual design report,
https:$//$slacportal.slac.stanford.\newline edu$/$sites$/$
lcls$\_$public$/$lcls$\_$ii$/$Published$\_$Documents$/$CDR$\%$20Index.pdf

\bibitem{MEAS}  G. Geloni, V. Kocharyan and E.~Saldin, "Ultrafast X-ray pulse measurement
method", http://arxiv.org/abs/1001.3544 , DESY 10-008 (2010).

\bibitem{PUMP} G. Geloni, V. Kocharyan and E.~Saldin, "Scheme for femtosecond-resolution pump-probe
experiments at XFELs with two-color ten GW-level X-ray pulses",
http://arxiv.org/abs/1001.3510 , DESY 10-004 (2010).

\bibitem{TAP1} A. Lin and J.M. Dawson, Phys. Rev. Lett. 42 1670 (1979)

\bibitem{TAP2} P. Sprangle, C.M. Tang and W.M. Manheimer, Phys. Rev. Lett. 43 1932 (1979)

\bibitem{TAP3} N.M. Kroll, P. Morton and M.N. Rosenbluth, IEEE J. Quantum Electron., QE-17, 1436 (1981)

\bibitem{TAP4} T.J. Orzechovski et al., Phys. Rev. Lett. 57, 2172 (1986)

\bibitem{FAWL} W. Fawley et al., NIM A 483 (2002) p 537

\bibitem{CORN} M. Cornacchia et al.,  J. Synchrotron rad. (2004) 11, 227-238

\bibitem{WANG} X. Wang et al., PRL 103, 154801 (2009)

\bibitem{S2ER} I. Zagorodnov, "Beam Dynamics Simulations for XFEL",
http://www.desy.de/xfel-beam/s2e (2011).

\bibitem{GENE} S Reiche et al., Nucl. Instr. and Meth. A 429, 243 (1999).



\end{thebibliography}
\end{document}